\documentclass[prd,twocolumn,groupedaddress,showpacs,floatfix,
nofootinbib]{revtex4}

\usepackage{graphicx}
\usepackage{latexsym}
\usepackage{amsmath}
\usepackage{amssymb}
\usepackage{amsfonts}
\usepackage{bm}
\usepackage{longtable}
\usepackage{dcolumn}
\usepackage[]{natbib}

\begin{document}

\title{Grid-based simulation program for gravitational wave interferometers 
\\ with realistically imperfect optics}

\author{Brett Bochner}
\email[Electronic addresses: ]{brett_bochner@alum.mit.edu, phybdb@hofstra.edu}
\affiliation{Department of Physics and Astronomy, Hofstra University, 
Hempstead, NY 11549}

\author{Yaron Hefetz}
\email{Yaron_Hefetz@hotmail.com}
\affiliation{14 Shoshanim St., Herzeliya, 46492, Israel}

\begin{abstract}

We describe an optical simulation program that models a complete,
 coupled-cavity interferometer like those used by the Laser Interferometer
 Gravitational-Wave Observatory (LIGO) Project. A wide variety of
 interferometer deformations can be modeled, including general surface
 roughness and substrate inhomogeneities, with no \textit{a priori}
 symmetry assumptions about the nature of interferometer imperfections.
 Several important interferometer parameters are optimized automatically
 to achieve the best possible sensitivity for each new set of perturbed
 mirrors. The simulation output data set includes the circulating powers and
 electric fields at various points in the interferometer, both for the main
 carrier beam and for its signal-sideband auxiliary beams, allowing an
 explicit calculation of the shot-noise-limited gravitational-wave
 sensitivity of the interferometric detector to be performed. Here we
 present an overview of the physics simulated by the program, and
 demonstrate its use with a series of runs showing the degradation of
 LIGO performance caused by realistically-deformed mirror profiles. We
 then estimate the effect of this performance degradation upon the
 detectability of astrophysical sources of gravitational waves. We conclude
 by describing applications of the simulation program to LIGO research
 and development efforts.
\end{abstract}


\maketitle

\section{\label{s1}Introduction}

The LIGO Project \cite{rf1} is part of the current initiative to detect
 gravitational radiation via its perturbing effects on resonant laser
 interferometers. This initiative involves several large collaborations
 around the world, including VIRGO \cite{rf2}, GEO \cite{rf3}, TAMA
 \cite{rf4}, and ACIGA \cite{rf5}.

A paramount issue in all such projects is the maximization of interferometer
 sensitivity to detect the extremely weak signals that are expected from even
 the most powerful astrophysical sources \cite{rf6}. To that end,
 coupled-cavity systems with multiple resonant stages will be used to
 maximize the shot-noise-limited signal-to-noise ratio of the gravitational
 wave (GW) signal readout. For these complex interferometric detectors,
 intensive modeling is necessary to estimate their performance in the presence
 of general optical imperfections.

A hierarchy of approaches has been used to estimate detector performance,
 each method negotiating the trade-off between accuracy and computational
 complexity. Analytical methods may suffice for the consideration of optical
 defects that can be treated as pure losses, or for some cases involving
 geometric \cite{rf7} or randomized \cite{rf8} mirror deformations. A matrix
 model that evaluates the coupling between the first few lowest-order TEM
 laser modes has been shown to be useful for the study of mirror tilts and
 beam displacements \cite{rf9}; and another matrix model using discrete
 Hankel transforms exists for problems with axial symmetry \cite{rf10}.
 Such models, which consider the exchange of power between a limited set of
 pre-specified modes, allow one to obtain fast results that can be used for
 predicting certain important interferometer behaviors (e.g., time-dependent
 detector responses \cite{rf11,rf12,rf13}), at the expense of some
 sophistication in modeling the detailed interferometer steady-state power
 buildup. For the consideration of highly general optical imperfections,
 however, the most comprehensive method is the complete modeling of the
 transverse structure of the laser field wavefronts; this method simulates
 the electric fields on large grids, and uses (Fourier transform-based)
 numerical computations for the propagations of these laser beams through
 long cavities \cite{rf14,rf15}. This technique is useful in the case of
 mirrors with complex deformations, and for mirrors with significant losses
 (due to absorption, scattering, and/or diffractive loss from finite-sized
 apertures), that violate the assumption of unitarity \cite{rf9} for mirror
 operators in the matrix models, thus introducing complications into that
 approach \cite{rf12}.

Grid-based simulations of intra-cavity laser fields have a long history
 (e.g., \cite{rf16}), and have been applied previously (e.g., 
 \cite{rf17,rf18,rf19}) to the study of the interferometric GW detectors now
 being implemented. But simplifications are generally imposed, such
 as restricting the optical deformations that are studied 
 (e.g., considering only geometric imperfections,
 like tilt and curvature mismatch \cite{rf17,rf18}), and/or by modeling a
 relatively simple cavity system \cite{rf19}. In this paper, we describe a
 simulation program (originally based upon the work of Vinet \textit{et al.}
 \cite{rf15}) that has been extended to efficiently model the complex fields
 that build up between realistically imperfect optical components, while
 resonating within complete, coupled-cavity interferometers like those used
 for the LIGO detectors. Versions of our program have been used for a variety
 of applications by the gravitational wave community, including numerous design
 and performance estimation tasks conducted by the LIGO group itself (see
 Section \ref{s6}), as well as for collaborative investigations between LIGO
 scientists and other groups \textemdash ~ such as ACIGA-LIGO efforts to 
 explore alternative interferometer length control schemes \cite{rf20}, and
 TAMA-LIGO efforts to estimate the effects of mirror imperfections
 \cite{rf21} and thermal lensing \cite{rf22} upon the performances of their
 future, large-scale interferometers.

A wide variety of interferometer imperfections can be modeled with our
 program, including mirror tilts and shifts, beam mismatch and/or
 misalignment, diffractive loss from finite mirror apertures, mirror
 surface figure and substrate inhomogeneity profiles \textemdash ~ 
 in particular,
 using deformation phase maps that are adapted from measurements of real
 mirror surfaces and substrates \textemdash ~ as well as fluctuations
 of reflection
 and transmission intensity across the mirror profiles. In addition to the
 main (carrier frequency) laser field, auxiliary fields (i.e., radio frequency
 sidebands) for LIGO's heterodyne signal detection scheme are also modeled,
 allowing us to give absolute numbers for the shot-noise-limited GW-sensitivity
 of a single LIGO interferometer. Furthermore, a number of active optimization
 procedures are performed continuously during code execution, guaranteeing that
 several key parameters in the LIGO configuration will be brought to their
 optimum values upon program completion; the GW-sensitivity is thus maximized
 for each specific ``realistic'' interferometer, given the particular
 imperfections being simulated in that run. Considering the time needed for
 program execution, we note that though our principal usage of the program has
 been on various supercomputing platforms, a full run of the simulation code
 (with all options included) can be performed in a non-prohibitive amount of
 time on a modest Sun SPARCstation, even for significantly (and
 non-symmetrically) deformed mirrors. This has been achieved through a
 combination of fast iteration techniques, efficient parameter optimization
 routines, and procedures designed to carefully choose the initial guesses for
 laser fields that are computed via relaxation. Lastly, we note that versions
 of the code are available for both the first-generation LIGO and advanced-LIGO
 (Dual Recycling \cite{rf23}) configurations \cite{rf28,newref1}, though we will
 focus primarily upon the former in this paper.

The discussion is organized as follows: in Section \ref{s2}, we give a
 description of the physical system that is modeled here (i.e., a
 first-generation LIGO interferometer), and the computed electric fields
 that are necessary for the calculation of its shot-noise-limited
 GW-sensitivity function. In Section \ref{s3}, we provide an overview of the
 technical details of the program's operations, including the modeling of the
 optics, the iterative method used for computing the interferometer's electric
 fields, and the various optimization procedures that are performed to maximize
 its sensitivity. In Section \ref{s4}, we present an array of runs with
 representative sets of mirror substrate and surface deformation maps adapted
 from real mirror measurements, in order to demonstrate how
 realistically-deformed mirrors will reduce the circulating power stored in a
 LIGO interferometer (as well as altering its modal structure), thus degrading
 the interferometer sensitivity and interfering with its control systems. In
 Section \ref{s5}, we discuss the impact of optical imperfections upon LIGO
 science capabilities, by estimating how deformed-mirror effects will reduce
 LIGO's ability to detect astrophysical sources of gravitational waves, such
 as non-axisymmetric pulsars and coalescing black hole binaries \cite{rf6}.
 In Section \ref{s6}, we conclude with a discussion of various LIGO research
 and development initiatives to which this grid-based simulation work has
 contributed.

\section{\label{s2}The Physical System that is Modeled}

\begin{figure}
\includegraphics{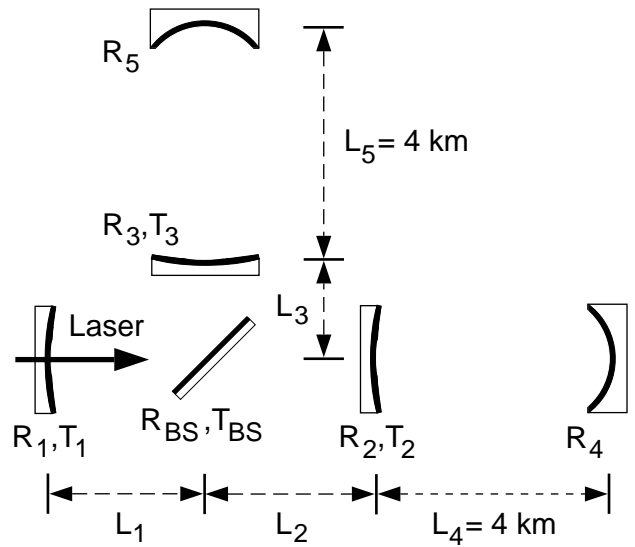}
\caption{\label{fg1}Schematic diagram of the core optical configuration of a
 first-generation LIGO interferometer. (Not drawn to scale.)}
\end{figure}

Figure \ref{fg1} is a schematic diagram of the core optical configuration of
 a full-sized, first-generation LIGO interferometer \cite{rf24}. A summary of
 typical interferometer (and computational) parameters for this configuration
 is presented in Table \ref{tbl1}. These are the primary program input values
 that we will use for the sample runs to be presented in Section \ref{s4}. Not
 shown (or modeled by us) are the mode cleaning, frequency stabilizing, and
 matching optics which prepare the laser light for the interferometer; also not
 modeled are the pickoffs, phase modulators, or the full control system
 apparatus that will be used in a real interferometer to read out its complete
 operational state \cite{rf25}.

\begin{table}
\caption{\label{tbl1}Typical parameter values for a first-generation LIGO
 interferometer, including physical specifications and computational parameters
 required for the simulation program. The labeling of the optical elements are
 as depicted in Fig.~\ref{fg1}. Some parameters are optimized during code
 execution, and are given here only as approximate ranges of values.}
\begin{ruledtabular}
\begin{tabular}{ll}
\textbf{Quantity} & \textbf{Value(s)}\\
\hline
\hline
Laser Wavelength & $1.064 ~ \mu \text{m}$ (Nd:YAG light)\\
\hline
Modulation Frequency & $\nu_{mod} \sim 24.0 ~ \text{MHz}$\\
\hline
 & $L_{1} = 5.0 ~ \text{m}$\\
 & $L_{2} = 4.19 ~ \text{m} + L_{\text{asymm}}$\\
Cavity Lengths & $L_{3} = 4.19 ~ \text{m} - L_{\text{asymm}}$\\
 & $L_{\text{asymm}} \sim 9-25 ~ \text{cm}$\\
 & $L_{4} = L_{5} = 4.0 ~ \text{km}$\\
\hline
 & $R_{\text{curv,1}} = 10.0 ~ \text{km}$\\
Mirror Curvature Radii & $R_{\text{curv,2}} = R_{\text{curv,3}} =
 14.6 ~ \text{km}$\\
 & $R_{\text{curv,4}} = R_{\text{curv,5}} = 7.4 ~ \text{km}$\\
\hline
 & $R_{1} \sim .9861-.9390$\\
Mirror Intensity Reflectivities & $R_{2} = R_{3} = .97$\\
(Refl.-Side) & $R_{4} = R_{5} = .99994$\\
 & $R_{\text{BS}} = .49992$\\
\hline
Mirror Intensity Reflectivities & $R_{1} \sim .9861-.9390$\\
(AR-Side) & $R_{2} = R_{3} = .968817$\\
 & $R_{\text{BS}} = .49971$\\
\hline
Mirror Intensity Transmissions & $T_{1} \sim .01385-.06095$\\
(Both Sides) & $T_{2} = T_{3} = .02995$\\
\textit{($\text{Pure Loss} \equiv 1 - R - T$)} & $T_{\text{BS}} = .50003$\\
\hline
Beam Waist Diameter & $7.0 ~ \text{cm}$\\
\hline
 & $24 ~ \text{cm}$ (Circular Profiles),\\
Mirror Aperture Diameters & $24.4 \times 17.2 ~ \text{cm}$ (Beamsplitter\\
 & at $45^{\circ}$ w.r.t. Beam Axis)\\
\hline
Mirror Thicknesses & Beamsplitter = $4 ~ \text{cm}$\\
(Perpendicular to Surface) & All Others = $10 ~ \text{cm}$\\
\hline
Substrate Refraction Index & $n = 1.44963$\\
\hline
Calculational Window Size & $70 ~ \text{cm} \times 70 ~ \text{cm}$ (Square)\\
\hline
Gridding of Calc. Window & $256 \times 256$ pixels\\
\end{tabular}
\end{ruledtabular}
\end{table}

The system depicted in Fig.~\ref{fg1}, essentially a Michelson interferometer,
 converts the differential arm length changes caused by a gravitational wave
 (GW) into an oscillating output field amplitude at the exit port of the
 beamsplitter, where a carrier field dark-fringe would otherwise (ideally)
 have been maintained. The partially-transmitting input mirrors
 $(T_{2}, T_{3})$ and highly-reflective end mirrors $(R_{4}, R_{5})$ form
 Fabry-Perot (FP) arm cavities in the ``inline'' and ``offline'' arms, which
 amplify this effect; and the power recycling mirror $(R_{1})$ provides a
 broadband amplification for the stored energy in the interferometer that
 is available for signal detection \cite{rf26}. Coupling occurs between the
 fields from each of the different cavities in this optical configuration,
 and is responsible for the complex response behaviors of the system.

Radio-frequency modulation sidebands (``RF-sidebands'') impressed upon the
 carrier-frequency input light serve as a ``local oscillator'' for a
 heterodyne detection scheme \cite{rf25}. Substantial RF-sideband power is
 made to emerge from the beamsplitter exit/signal port, and it is added to
 the quantity of GW-induced carrier light that is escaping there, to
 ultimately produce an output signal linear in \textit{h}, the
 dimensionless amplitude of the GW. The carrier power is maximized by
 giving it a double-resonance: it is first resonant in (either of)
 the FP arm cavities, and again resonant in the ``power recycling
 cavity'' (PRC) formed by mirrors $R_{1}$, $R_{2}$, and $R_{3}$. Alternatively,
 the RF-sidebands (in this standard LIGO detection scheme) are only resonant in
 the PRC, to take advantage of the broadband amplification there; but they are
 far-off-resonance in the long FP-arms, so that they may serve as a stable
 reference which is unaffected by gravitational waves.

Assuming that the carrier is held to its double resonance (giving it a phase
 shift of $\pi$ in resonant reflection from the FP-arms \cite{rf24}), the
 RF-sidebands have their own resonance requirements, defined by the two
 conditions: $(2 ~ k_{\text{mod}} ~ L_{\text{arm}} ) \approx N_{\text{odd}}
 \cdot \pi$, and $(2 ~ k_{\text{mod}} ~ L_{\text{PRC}}) \approx N_{\text{odd}}
 \cdot \pi$. Referring to Fig.~\ref{fg1}, $L_{\text{arm}}$ represents $L_{4}$
 or $L_{5}$, $L_{\text{PRC}} \equiv L_{1} + (L_{2} + L_{3})/2$, and the
 RF-modulation frequency is defined according to $k_{\text{mod}} = 2 \pi
 \nu_{\text{mod}}/c \equiv 2 \pi \vert \nu_{\text{carr}} -
 \nu_{\text{SB}}\vert /c$. The first (anti-resonance) condition is not
 enforced as a precise equality, in order to avoid the unwanted resonance
 of non-negligible second-order modulation sidebands
 (at $2 ~ \nu_{\text{mod}}$) in the FP arm cavities. But the second
 (PRC-resonance) condition must be achieved to sub-wavelength tolerances,
 and requires careful fine-tuning corrections due to the effects of
 realistically-deformed optics (see Sec.~\ref{s3s3s2}).

The heterodyne GW-signal is obtained by interfering the GW-induced carrier
 beam at the beamsplitter exit/signal port with the emerging sideband beams,
 and demodulating the resultant photodiode output signal at
 $\nu_{\text{mod}}$. Efficient coupling of the RF-sidebands to the signal
 port is achieved by incorporating a macroscopic length asymmetry between
 the two arms of the PRC, $L_{\text{asymm}} \equiv (L_{2} - L_{3})/2$,
 thus allowing an optimal fraction of sideband power to be extracted at the
 exit port during each round-trip through the PRC, even while the carrier
 is held to a dark-fringe there. This signal generation method (see
 Sec.~\ref{s3s3s4}) is referred to as the ``Schnupp asymmetry
 scheme'' \cite{rf27}.

To simplify our simulation task, we model only those aspects of the LIGO
 system which have a direct role in generating the GW-signals: the
 aforementioned carrier beam, and its RF-sidebands, resonating in the core
 optical system of Fig.~\ref{fg1}. If we make the (good) approximation that 
 most of the sensitivity-limiting shot noise at the beamsplitter exit port 
 is due to these fields \textemdash ~ some carrier field power emerging
 (primarily) because of imperfect dark-fringe contrast, and
 RF-sideband field power being maximally channeled to the exit port
 \textemdash ~ and neglect contributions from various other
 fields used for interferometer control systems, or from other (controlled)
 noise couplings \cite{rf25}, then we can do a full calculation of the
 shot-noise-limited GW-sensitivity of the simulated interferometer. Beyond
 this, rather than explicitly emulating all of the control systems of the
 real LIGO, the simulation program uses alternative methods (see
 Sec.~\ref{s3s3}) for the numerous parameter adjustments (resonance
 finding, etc.) that are needed to optimize interferometer performance;
 and it does so in a manner that reflects the behavior of the real system
 as accurately as possible.

In summary, the primary goal of our modeling efforts is to completely solve
 for the steady-state carrier and sideband fields that resonate in a LIGO
 interferometer with realistic optical imperfections, which is held to its
 proper operating point, and is optimally configured for maximal
 GW-sensitivity. The full effects of optical imperfections upon the
 sensitivity of interferometric detectors are then determined from these
 simulated cavity fields.

\section{\label{s3}The Fundamentals of the Simulation Process}

Here we present an overview of the program and its operations during execution,
 and the additional tasks which must be done during pre- and post-processing
 stages in the course of performing simulation experiments. The full technical
 details of our work can be found in \cite{rf28}.

In regards to computing platforms, the simulation program was originally
 written in the SPARCcompiler 3.0 version of Fortran 77. A complete run
 (including carrier and sideband frequencies, with all interferometer fields
 computed and all optimizations done) with a set of non-ideal mirrors, and
 with (for example) $128 \times 128$ pixelized grid maps used for the optical
 computations, takes less than a day on a 2-processor SPARCstation 20. This
 computation time is reduced to a couple of minutes when the program is
 executed on massively-parallel supercomputing platforms
 (e.g., \cite{rf29}\footnote{We are grateful to T. Phung and H. Lorenz-Wirzba of
 the Center for Advanced Computing Research (CACR) at the California
 Institute of Technology, for their extensive help in adapting our program
 (both the first-generation LIGO and the advanced-LIGO Dual Recycling code
 versions) to its initial, massively-parallel supercomputing platform, the
 CACR Paragon Intel i860 clusters \textit{Trex} and \textit{Raptor}.},
 \cite{rf30}\footnote{We are grateful to E. D'Ambrosio, R. Jenet,
 G. Jennings, B. Keig, B. Kells, A. Weinstein, and S. Wiley for their work
 in adapting our program (both the first-generation LIGO and the
 advanced-LIGO versions) for use on the CACR V-Class HP-UNIX cluster
 (skinner.cacr.caltech.edu) at the California Institute of Technology,
 and for using it to conduct extensive computational investigations on
 that massively-parallel computer platform.}). The runs to be presented
 in this paper were performed using 32 nodes of the Paragon machine
 \textit{Trex} \cite{rf29}, a 512 (compute) node machine utilizing Intel
 i860 processors.

\subsection{\label{s3s1}Basic Optical Operations}

\begin{figure}
\includegraphics{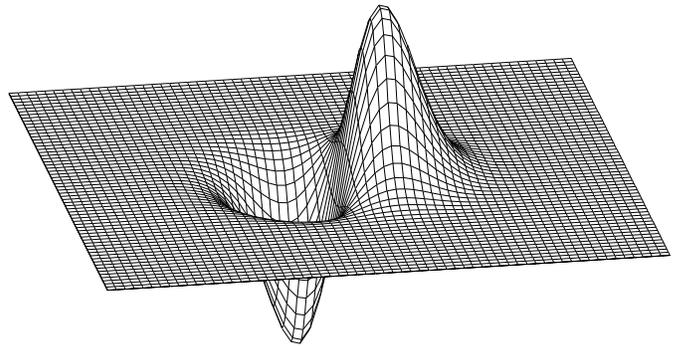}
\caption{\label{fg2}Transverse slice of a Hermite-Gaussian TEM$_{10}$ mode
 (only the real part is shown), taken at the waist plane of the beam, and
 recorded on a $64 \times 64$ pixelized grid.}
\end{figure}

We use the customary approach \cite{rf15,rf17,rf18,rf19,rf20,rf21,rf22} for
 the grid-based modeling of the laser field wavefronts: a primary propagation
 direction is assumed for the collimated beam(s) in each part of the
 interferometer, and a perpendicular slice can be taken anywhere along the
 beam propagation axis, at locations of interest. Each of these beam slices
 is recorded on a two-dimensional (2D) grid, with a pixel entry representing
 the complex electric field (``e-field'') amplitude at that transverse spatial
 position in the slice. Figure \ref{fg2} is an example of such a grid-based
 electric field, in particular that of a Hermite-Gaussian TEM$_{10}$ mode
 \cite{rf14}. No polarization vector is currently recorded in our grids
 (we cannot, for example, model birefringence effects). The precision that
 can be achieved by the program depends upon how accurately it simulates the
 two basic physical processes which must be performed on the interferometer
 e-fields: \textit{propagations}, and \textit{interactions with mirrors}. We
 discuss propagations first.

The program utilizes Siegman's method \cite{rf14} for the plane-to-plane
 propagation of light in the paraxial approximation, performed via a three
 step process: a Fourier transform and an inverse transform, sandwiched
 around a pixel-by-pixel multiplication step in spatial-frequency-space
 (``\textit{k}-space''), in which the e-field slice is multiplied with a
 distance-dependent 2D propagator matrix of the same pixelization. For each
 curved, potentially jagged mirror profile to be simulated, a flat plane is
 defined near that side of the optic to serve as a reference plane.
 Propagations thus translate an e-field slice through the large distance
 from an initial reference plane to a destination reference plane near a
 mirror (or at any chosen position along the beam axis). The computationally
 intensive parts of this process (i.e., the transforms) can be performed
 rapidly by a Fast Fourier Transform (FFT) routine (e.g., \cite{rf31}). All
 such macroscopic-distance propagations (i.e.,
 $\Delta L \gg \lambda_{\text{carr}}$) of e-fields are done in this manner.

A very important issue to deal with for propagations is the problem of
 \textit{aliasing} \cite{rf31}, a common complication for calculational
  procedures that use discrete Fourier transform methods. Aliasing can create
  artifacts in the simulation results if (relatively) large-angle scattering
  sends power beyond the edge of the grid calculational window during the
  propagations. Such power automatically re-enters the calculation from the
  other side of the grid, and may (fraudulently) be incorporated back in the
  physical simulation, instead of being filtered out, as it would be by
  absorbing baffles in a real interferometer. This problem is most significant
  for the long propagations through the 4 km FP arm cavities.

We can reduce (or eliminate) such aliasing by filtering the relevant
 \textit{k}-space matrix that will multiply an e-field slice, in between the
 Fourier transform and inverse-transform steps of a propagation. The goal is
 to preserve as much ``real power'', while eliminating as much ``aliased
 power'', as possible; the distinction between them is that the former always
 stays within the apertures of the finite-sized mirrors, while the latter
 leaves the grid completely, but re-enters from the other side and comes far
 enough into the middle of the grid to fall once again within the mirror
 apertures.

Let $W$ be the physical side-length of the square calculational window,
 $A$ be the mirror aperture diameter, and $L$ be the propagation distance
  length. The cutoff between physically real power and ``aliasing'' power
  is a matter of propagation angle: power traveling at
  $\theta < \theta_{\text{r}} \equiv A/L$ may be able to stay within
  the mirrors, while power with $\theta > \theta_{\text{a}} \equiv (W-A)/L$
  usually cannot, but will sometimes be able to leave the grid \textit{and}
  return to erroneously re-enter the mirror apertures. For a discrete grid
  with 
  $N_{\text{pix}}$ pixels on a side (numbered $n = -(N_{\text{pix}}/2 - 1)
  \dots (N_{\text{pix}}/2)$), these angles correspond to (respectively) the
  pixel numbers $N_{\text{r}} = \text{Int}[A \cdot W/(L \cdot \lambda)]$ and
  $N_{\text{a}} = \text{Int}[(W - A) \cdot W/(L \cdot \lambda)]$, where
  $\lambda$ is the light wavelength. All of the aliasing power can be safely
  eliminated by nulling pixels with $\vert n \vert > N_{\text{a}}$ in the
  \textit{k}-space propagator matrix, without removing any real power, as long
  as $N_{\text{a}} > N_{\text{r}}$. But if $N_{\text{a}} < N_{\text{r}}$, then
  one must choose some compromise between \textit{keeping} real power and
  \textit{cutting out} aliasing power for pixels $N_{\text{a}} < \vert n \vert
  < N_{\text{r}}$. One can, of course, force $N_{\text{a}} > N_{\text{r}}$ to
  be true by increasing the calculational window size to be much larger than
  the mirrors (i.e., $W > 2A$), but this requires more pixels to be used in the
  grid (e.g., $256 \times 256$), which increases the computational load.
  Making $W$ large without using an adequate number of pixels would lead to
  poor sampling of the laser beam itself \cite{rf28}, thus creating a new
  aliasing problem, due to the inadequate resolution of the discrete grids.
  In our runs, we use large calculational windows and many pixels whenever
  possible (such as for the runs presented below); but when necessary, the
  program has an option which applies an apodization scheme to ``trim'' the
  propagation operators in a way that enforces a graduated compromise between
  keeping real power and eliminating aliasing power.

Next, we consider mirror interactions. The program must carry out two basic
 mirror interaction operations: reflections ($r$) and transmissions ($t$).
 Mirrors will not be perfectly uniform or flat in the direction transverse to
 the beam; they will have inhomogeneities in the refraction index and/or
 thickness of their substrates, spatially-varying surface height profiles,
 variations in the quality of the reflective coatings, macroscopic curvatures,
 tilts, etc. This leads to optical path length variations (i.e.,
 spatially-varying phase delays) across their profiles, as well as variations
 in the \textit{amplitudes} of $r$ and $t$, the latter mostly due to variations
 in the reflective side or anti-reflective (AR) side coatings. These phase and
 amplitude effects are simulated by creating complex mirror maps for all $r$
 and $t$ operators, which are also recorded on 2D pixelized grids. As
 demonstrated by Vinet \textit{et al.} \cite{rf15}, it is a good (``short
 distance'') approximation to treat each pixel of the beam as an independent
 little plane wave, and reflect (or transmit) that piece of the e-field by
 multiplying that pixel in the e-field map by the corresponding pixel in the
 relevant mirror map, so that each e-field pixel interacts only with the mirror
 pixel located immediately in front of it. Thus each mirror reflection or
 transmission operation is reduced to a pixel-by-pixel multiplication step
 between a mirror operator map and the e-field slice on the reference plane
 near to it. The simulation program uses these mirror interaction operations,
 in combination with the propagation algorithm described above, to model the
 entire behavior of the laser fields in the interferometer.

There are some limitations in using this methodology for mirror interactions,
 besides the obvious one of restricting it to operations over short distances
 (i.e., in the near-field). Vinet \textit{et al.} \cite{rf15} derive a
 requirement which specifies how small the deviations of a mirror profile can
 be from its idealized shape, before this pixel-by-pixel multiplication method
 loses a large amount of accuracy compared to the exact calculation via the
 Huygens-Fresnel integral formulation in scalar wave theory \cite{rf14}.
 Realistically imperfect (LIGO-quality) mirrors easily satisfy this requirement.
 Furthermore, as noted by Tridgell \textit{et al.} \cite{rf17}, the difference
 in phase between one pixel in a mirror map and its neighbor must be smaller
 than $2 \pi$ if a continuously-varying surface displacement on the mirror is
 to be adequately sampled by the grid. For our simulation parameters (cf.~Table
 \ref{tbl1}), this limits mirror tilts to $\theta < 10^{-4}$ radians (not
 including the base $45^{\circ}$ tilt of the beamsplitter, which is handled
 separately), and mirror curvature radii to $R_{\text{curv}} > 1.3 ~ \text{km}$;
 both of these limitations are easily satisfied in our runs. Finally, we
 supplement these requirements with a general rule-of-thumb: a \textit{tiny
 Gaussian beam}, with a waist size equal to the width of a pixel (\textit{and}
 an initial propagation direction with respect to the beam axis that is defined
 by the e-field's wavefront curvature at the location of that pixel), must not
 expand or shift over so much so that it would relocate a significant fraction
 of its power onto any neighboring pixels, during the entire course of the
 mirror interaction. If that rule is broken, then this method for mirror
 interactions will be inaccurate, and the pixel-by-pixel multiplication of
 maps will not be sufficient for modeling reflection or transmission
 operations.

To create a complete (but not over-determined) description of a mirror, we
 consider the full information expressed by mirror interactions, as well as
 physical symmetries and the conservation of energy. A 2-port mirror
 (reflective- and AR-sides) must relate 2 complex input fields to 2 complex
 output fields. Thus 4 complex (or 8 real) elements are needed, at each pixel
 location, to specify that pixel of mirror completely. These 4 complex elements
 correspond to the complex reflection and transmission operations performed
 from the two different sides of the mirror. The two transmission operations
 from either side \textemdash ~ barring excessive beam expansion or
 focusing during the
 transmission \textemdash ~ must in fact be the same
 \cite{rf32}: $t_{\text{right}} = t_{\text{left}} \equiv t$
 (except for the beamsplitter, which requires distinct transmission maps for
 the inline and offline paths). The mirror description is thus reduced to
 3 complex elements per pixel, i.e., 3 complex mirror maps: the transmission
 map \textit{t}, and reflection maps from either side, $r_{\text{refl.}}$ and
 $r_{\text{AR}}$.

Next, by energy conservation we have (for each mirror pixel):
 $1 - \vert t \vert^{2} - \vert r_{\text{refl.}} \vert^{2}
 \equiv A_{\text{refl.}} \geq 0$, and $1 - \vert t \vert^{2} -
 \vert r_{\text{AR}} \vert^{2} \equiv A_{\text{AR}} \geq 0$,
 where $A_{\text{refl.}}$, $A_{\text{AR}}$ are the losses experienced in
 reflection from either side of the mirror. But these conditions are not
 sufficient. By considering the complete, superposed e-fields which exist
 on either side of the mirror, both before and after the mirror interaction
 (i.e., the incoming e-fields vs. the outgoing ones), and by requiring that
 the total power in all e-fields \textit{must not increase} due to the
 interaction, one may obtain the following inequality (expressed in terms
 of the incoming, ``before'' fields):
\begin{widetext}
\begin{equation}
P^{\text{before}}-P^{\text{after}}=A_{\text{refl.}} \cdot
 |E^{\text{before}}_{\text{refl.-side}}|^{2} +A_{\text{AR}}
 \cdot |E^{\text{before}}_{\text{AR-side}}|^{2} -2\Re\{(t^{\ast}
 \cdot r_{\text{refl.}} + t \cdot r_{\text{AR}}^{\ast})
 \cdot E^{\text{before}}_{\text{refl.-side}}
 \cdot (E^{\text{before}}_{\text{AR-side}})^{\ast} \} \geq 0~.
\label{dispeq1}
\end{equation}
\end{widetext}

For the simplest case of a loss-free mirror ($A_{\text{refl.}}
 = A_{\text{AR}}$ = 0, and $P^{\text{before}} - P^{\text{after}} = 0$),
 this just reduces to the complex generalization of one of Stokes relations
 \cite{rf32}: $(r_{\text{refl.}}/r_{\text{AR}}^{\ast}) = -(t/t^{\ast})$.
 Assuming the transmission coefficient $t$ to be real, this generates the
 familiar possibilities for the phase relationships between the reflectivities
 on either side of a mirror: e.g., $(r_{\text{refl.}}, r_{\text{AR}})
 \propto (i,i)$, $(r_{\text{refl.}}, r_{\text{AR}}) \propto (\pm 1, \mp1)$,
 etc. For the more general case of a lossy mirror, we can convert
 Eq.~(\ref{dispeq1}) into a simpler prescription by considering variations
 to the relative phases and amplitudes of
 $E^{\text{before}}_{\text{refl.-side}}$
 and $E^{\text{before}}_{\text{AR-side}}$, thus generating the strict energy
 conservation condition:
\begin{equation}
\frac{\sqrt{A_{\text{refl.}} \cdot A_{\text{AR}}}}{|t^{\ast} \cdot
 r_{\text{refl.}} + t \cdot r_{\text{AR}}^{\ast}|} \geq 1~.
\label{dispeq2}
\end{equation}
This inequality (generalized further for the beamsplitter) must be satisfied
 (at every pixel) to guarantee conservation of energy in a mirror, given
 any e-fields which could be incident upon it. Our simulation program enforces
 this condition by testing the input data maps for each mirror, and rejecting
 runs with unphysical specifications.

Given all of these requirements, the pixelized mirror maps described above
 are versatile tools for modeling a wide variety of optical features and
 imperfections, as discussed in Sec.~\ref{s1}.

\subsection{\label{s3s2}Relaxation of Steady-State Electric Fields}

Our program models a static interferometer, assumed to be held to the
 correct operating point for an indefinite period of time. This assumption
 enormously simplifies the program, and our overall simulation task, while
 still enabling us (see Sec.~\ref{s3s4}) to compute the frequency-dependent
 GW-response of a LIGO interferometer. (The dynamics of LIGO control systems
 do not necessarily come into this calculation, since the mirrors act like
 ``free-masses'' at the GW-frequencies that LIGO is most sensitive to, and
 because the GW's are presumably not strong enough to interfere with
 the interferometer's resonant lock.) The primary task of the simulation
 program, therefore, is to
 compute the relaxed, steady-state, resonant electric fields that build up in
 each part of the interferometer, when it is excited by a laser beam of fixed
 amplitude, orientation, and frequency (at the nominal carrier and sideband
 frequencies), entering through the power recycling mirror.

The multiply-coupled-cavity nature of the interferometer, and the physical
 complexity of the simulation, means that the program must iteratively solve
 for an irreducible number of resonant e-fields (one per cavity) in the
 interferometer. For the first-generation LIGO configuration, three e-fields
 must be computed via relaxation: one in the PRC, and one in each of the FP
 arm cavities (though the precise location of each relaxed e-field in its
 cavity can be arbitrarily chosen). When the relaxation algorithm is finished,
 these three principal e-fields can be propagated, reflected, transmitted,
 and/or superposed together in order to generate the complete steady-state
 e-fields everywhere in the system.

For each e-field that will be relaxed, there is a steady-state equation of
 the form:
\begin{equation}
\roarrow{E}_{\text{steady-state}} = \Hat{\text{A}}
 \{\roarrow{E}_{\text{steady-state}}\} + \roarrow{E}_{\text{exc}}~.
\label{dispeq3}
\end{equation}
The left hand side of Eq.~(\ref{dispeq3}) is the e-field to be solved for;
 we display it here as a vector because it has a propagation direction:
 either ``forward'' or ``backward'' along the beam axis. The expression,
 $\Hat{\text{A}} \{\roarrow{E}_{\text{steady-state}}\}$, represents this
 steady-state e-field after it has gone on a round-trip through the
 interferometer, and has returned to its starting point. (Note that this
 ``round-trip'' includes all possible closed-loop paths through the
 interferometer that do not ever pass through the same
 location \textemdash ~ with the same propagation
 direction \textemdash ~ of any of the other principal e-fields
 being relaxed.) Lastly, $\roarrow{E}_{\text{exc}}$ represents a
 composite excitation e-field, which may consist of both a primary
 excitation e-field (e.g., the input laser beam, for the e-field being
 relaxed in the PRC), plus any ``leak'' e-fields that arrive from the
 other principal fields being relaxed (e.g., fields leaking from the
 FP-arms into the PRC). It is this leak-field part of the excitation term
 which is the source of many coupled-cavity effects in the interferometer;
 and it is the presence of the $\Hat{\text{A}}
 \{\roarrow{E}_{\text{steady-state}}\}$ term in Eq.~(\ref{dispeq3}) which
 makes $\roarrow{E}_{\text{steady-state}}$ ``self-coupled'', thus requiring
 us \cite{rf28} to use an iterative process to solve for it.

Specifying a relaxation convergence scheme is done by prescribing how the
 $(N+1)^{\text{th}}$ iterative guess for a given e-field is obtained from
 its $N^{\text{th}}$ iteration (and from the $N^{\text{th}}$ iteration of the
 other e-fields being relaxed simultaneously). Considering Eq.~(\ref{dispeq3}),
 the simplest possibility is to make the choice:
\begin{equation}
\roarrow{E}^{(N+1)} = \Hat{\text{A}} \{\roarrow{E}^{(N)}\} +
 \roarrow{E}^{(N)}_{\text{exc}} ~.
\label{dispeq4}
\end{equation}
This process can then be repeated for as many iterations as necessary, until
 the steady-state equation, Eq.~(\ref{dispeq3}), is satisfied by
 $\roarrow{E}^{(N)}$ to within a pre-specified threshold of accuracy
 (typically 1 part in $10^{4}$ in power, for our runs). This relaxation
 formula is guaranteed to succeed at smoothly converging an e-field to its
 correct steady-state form \textemdash ~ barring complications which
 may arise from
 changes to the interferometer caused by the parameter optimization routines
 (see Sec.~\ref{s3s3}) \textemdash ~ because it imitates how power
 actually does build up,
 through a sum of many bounces, in the cavity system of a real interferometer.
 This iteration process has been used successfully by previous researchers
 (e.g., \cite{rf15}).

Since this method does model the true physical buildup of power, however,
 it requires a great many iterations to converge, especially for coupled-cavity
 systems with large Q-factors (i.e., long transient decay times). We have found
 this relaxation scheme to be forbiddingly slow for a full-LIGO simulation
 program.

Therefore, as first suggested (and implemented, for a simpler cavity
 arrangement) by a LIGO colleague \cite{rf33}, our simulation program uses
 a different approach. Instead of Eq.~(\ref{dispeq4}) for choosing the
 $(N+1)^{\text{th}}$ iteration, we use the following expression:
\begin{equation}
\roarrow{E}^{(N+1)} = a \cdot \roarrow{E}^{(N)} + b
 \cdot (\Hat{\text{A}} \{\roarrow{E}^{(N)}\} +
 \roarrow{E}^{(N)}_{\text{exc}}) + c \cdot \roarrow{E}^{(N)}_{\text{exc}} ~,
\label{dispeq5}
\end{equation}
where $(a,b,c)$ are unknown, complex coefficients that are solved for by
 minimizing the error in what the steady-state equation will become in the
 \textit{next} round, with $\roarrow{E}^{(N+1)}$ (as a function of these
 unknown coefficients) taking the place of
 $\roarrow{E}_{\text{steady-state}}$ in Eq.~(\ref{dispeq3}),
 and $\roarrow{E}^{(N)}_{\text{exc}}$ taking the place of
 $\roarrow{E}_{\text{exc}}$ (noting that $\roarrow{E}^{(N)}_{\text{exc}}$
 does not yet take into account the new
 $(a,b,c)$ coefficients that will be chosen for the \textit{other}
 e-fields being iterated). The resulting expression for the steady-state
 error can be differentiated with respect to the six available degrees of
 freedom (the real and imaginary parts of $a$, $b$, and $c$), resulting in
 six simultaneous equations that are solved via matrix inversion. This
 relaxation method (which essentially reverts to the simpler method if we
 set $(a,b,c) = (0,1,0)$) gains a huge advantage compared to the method of
 Eq.~(\ref{dispeq4}), by having these additional, useful degrees of freedom
 available for each iteration \cite{rf33}. The number of iterations
 necessary to achieve convergence are greatly reduced (often by $\sim$1-2
 orders of magnitude), resulting in much faster e-field relaxation.

This ``abc'' iteration algorithm does have certain drawbacks. It is more
 difficult to implement in code, and it requires more information to be
 stored (thus using more memory), and more propagations to be performed
 (despite what is stated in \cite{rf33}, because we use three unknown
 coefficients instead of two: $c \neq 0$), during each iteration. It is also
 somewhat less stable, because of the wide range of abc-coefficients that
 may be chosen by the error minimization algorithm; in fact, large excursions
 in the iterated e-field structure and power level appear to be
 \textit{required} during (typically) the first $\sim$50-100 iterations,
 in order for the accelerated convergence scheme to function properly. We
 have found that the best (straightforward) way to enhance the convergence
 stability of the ``abc'' relaxation algorithm, without slowing it down to
 the pace of the method of Eq.~(\ref{dispeq4}), is to place hard limits upon
 the allowed choices of the abc-coefficients during the early stages (the
 first $\sim$100-200 iterations) of runs. This allows the relaxation process
 to avoid failure during the initial, large e-field adjustments, after which
 it settles down to converge efficiently to the steady-state solution.

Lastly, we note that the choice of $(a,b,c)$ for any one field will affect
 future iterations of the other fields, because they are all coupled via
 the $\roarrow{E}_{\text{exc}}$ leak-fields. Iterating them independently from
 one another ignores this coupling, and causes some slowing (and oscillation)
 of the relaxation process (we see a temporary ``sloshing'' of power back and
 forth between the PRC field and the FP arm cavity fields). This problem can
 be eliminated by choosing the $(a,b,c)$ coefficients for all three relaxed
 e-fields \textit{simultaneously}, by calculating the 18 unknown parameters
 (i.e., 9 complex ``abc'' coefficients) in order to minimize a
 specially-weighted sum of the iteration errors for all three relaxed fields.
 Such a ``global abc'' relaxation scheme is significantly more demanding in
 terms of coding difficulty and computer memory usage, but it is capable of
 making the relaxation process more stable, and further reduces the number of
 iterations needed to reach convergence. Though we have not yet implemented
 this global relaxation scheme into simulations of the first-generation LIGO
 interferometer, we have successfully incorporated global abc relaxation
 (with four relaxed e-fields needed, instead of three) into the version of
 our code used for the advanced-LIGO Dual Recycling configuration,
 with good results \cite{rf28}.

In summary, the ``abc'' iteration method presented here is a sufficiently
 reliable and \textit{extremely fast} relaxation scheme which greatly reduces
 program execution time, thus enabling us to perform full-LIGO simulation runs
 (with significant optical deformations) in a reasonable amount of time.

\subsection{\label{s3s3}Parameter Optimizations}

At the core of our efforts to make a realistic simulation of a LIGO
 interferometer are several procedures which bring the system into an
 ``optimal'' configuration for signal detection. The problem of optimization
 is a highly nontrivial matter: not only does the interferometer possess
 numerous degrees of freedom that must be optimized (e.g., all resonant cavity
 lengths, the Schnupp length asymmetry, etc.), but each evaluation of
 performance (i.e., GW-sensitivity) as a function of these optimizable
 parameters is extremely time consuming, since it requires the detailed
 computation of the carrier and sideband e-fields everywhere in the
 interferometer. It therefore appears infeasible to use the brute-force
 method of optimizing the interferometer's GW-sensitivity function by
 evaluating it for a thorough sampling of points over the entire,
 multidimensional parameter space.

As an alternative, each key parameter is optimized separately in our program,
 using some error signal (to be brought to zero), or some function of merit
 (to optimize), which is strongly (and solely) dependent upon that individual
 parameter, and which is a true measure of when that parameter is well chosen
 for the maximization of GW-sensitivity. This strategy is aided by the fact
 that the interferometer's GW-sensitivity is quite insensitive to particular
 \textit{combinations} of parameter changes, such that if certain optimizable
 parameters are displaced from one apparently optimal point in the
 multidimensional parameter space, then the other parameters will adjust
 themselves (via our optimization procedures) to compensate with virtually
 no reduction in overall interferometer sensitivity. We note that these
 optimization routines in our program are performed concurrently with the
 e-field relaxations, so that the final results emerging from the iterative
 relaxation scheme are the steady-state e-fields of a \textit{fully-optimized}
 interferometer.

The following subsections give an overview of the parameters that the program
 optimizes, the criteria for optimizing them, and the physical considerations
 which underlie their significance. Note that we have performed extensive
 empirical tests to verify that each of the optimization procedures discussed
 below works as desired.

\subsubsection{\label{s3s3s1}Length Adjustments for Carrier Resonance}

As discussed in Sec.~\ref{s2}, the carrier frequency beam must have a double
 resonance in the system consisting of the power recycling cavity and
 Fabry-Perot arm cavities, while being held to a dark-fringe at the exit port
 of the beamsplitter. Our method for achieving these resonance (and
 dark-fringe) conditions is similar to that of Vinet \textit{et al.}
 \cite{rf15}, in which we null the computed ``phases'' between certain
 specified cavity e-fields by adjusting the various cavity lengths. (A length
 change will alter the phase relationship between an e-field that has taken
 a round-trip through an adjusted path length, and one that has
 not \textemdash ~ potentially bringing a cavity to resonance.)
 Alternatively, we have
 chosen not to use the method of McClelland \textit{et al.} \cite{rf18},
 in which the e-fields are re-relaxed for each trial set of lengths until
 the configuration for maximum power buildup is found, since that would
 involve a time-consuming search over a multi-dimensional phase space of
 independent, controllable cavity lengths.

The phase $\Phi$ between two e-fields is defined via an overlap integral
 (or more precisely, by a discrete sum over the pixelized $N \times N$ grids),
 as follows:
\begin{widetext}
\begin{eqnarray}
\Phi[\roarrow{E}_{1},\roarrow{E}_{2}] \equiv \text{tan}^{-1}[\frac{
 \Im <\roarrow{E}_{1}|\roarrow{E}_{2}>}{ \Re
 <\roarrow{E}_{1}|\roarrow{E}_{2}>}]~, \nonumber \\ 
 \text{with:}~ <\roarrow{E}_{1}|\roarrow{E}_{2}> \equiv
 \frac{(\text{Calc. Window size})^{2}}{N^{2}} \cdot \sum_{i=1}^{N}
 ~ \sum_{j=1}^{N} & \roarrow{E}_{1}^\ast(i,j) \cdot \roarrow{E}_{2}(i,j) ~.
\label{dispeq6}
\end{eqnarray}
\end{widetext}
The phase between an e-field and its round-tripped analogue can be driven to
 zero by a (sub-wavelength) path length change, computed via the formula:
 $\Delta L = - \Phi \cdot (-\lambda_{\text{laser}} / 4 \pi) \equiv \Phi /
 2 k $ (an extra factor of $1/2$ is included here because the round-trip
 doubles the phase change caused by a cavity length adjustment). Note that
 this procedure cannot simply be performed once \textemdash ~ these
 phases depend in
 detail upon the structure of the iterated e-fields, and they must be
 repeatedly measured and adjusted throughout the e-field relaxation process.

To achieve resonance in a particular cavity, the proper solution is to ensure
 that the fields $\roarrow{E}_{\text{steady-state}}$,
 $\Hat{\text{A}} \{\roarrow{E}_{\text{steady-state}}\}$,
 and $\roarrow{E}_{\text{exc}}$ (as defined in Sec.~\ref{s3s2}) all have
 zero phase between them. (This condition of mutually zero phase is given
 some theoretical justification in the literature \cite{rf33}, as well as
 being apparent from Eq.~(\ref{dispeq3})). The phase-nulling procedure is
 performed in the FP arm cavities and in the PRC by performing microscopic
 adjustments to the three relevant cavity lengths here: $L_{4}$, $L_{5}$,
 and the ``common-mode'' recycling cavity length, $L_{\text{PRC}}$. Similarly,
 the dark-fringe condition at the beamsplitter exit port is achieved by setting
 the phase between the carrier e-fields coming from the inline/offline
 recycling cavity arms to an odd multiple of $\pi$, by microscopically
 adjusting $L_{\text{asymm}} \equiv (L_{2} - L_{3})/2$ via
 ``differential-mode'' corrections to $L_{2}$ and $L_{3}$.

Unlike the procedure discussed in \cite{rf15}, we do not choose any particular
 spatial mode (such as the lowest order, TEM$_{00}$ mode) of the interferometer
 e-fields for calculating these phases. Rather, we use the entire e-fields, for
 two reasons: first, for the dark-fringe condition, one wishes to minimize the
 \textit{total} dark-port power emerging from the beamsplitter exit port, all
 of which contributes to the shot noise; second, for the carrier resonance
 conditions, coupling between modes brings power back into the TEM$_{00}$ mode
 from higher modes, so that bringing the \textit{total} field (i.e., the
 ``perturbed interferometer mode'') to resonance \textit{also} maximizes the
 TEM$_{00}$ power used in calculations of the GW-signal. In any case, we have
 observed that these two methods (i.e., with or without spatial mode selection
 for resonance-finding) typically give very similar results.

\subsubsection{\label{s3s3s2}Sideband Frequency Fine-Tuning}

The RF-sidebands must also satisfy important conditions, specifically resonance
 in the PRC, and near-anti-resonance in the FP arm cavities. These conditions
 are affected by the optics, and the sidebands must also therefore be tuned for
 optimum performance.

As a first approximation, the cavity lengths and RF-modulation frequency are
 initialized through an analytic evaluation designed to simultaneously optimize
 carrier and RF-sideband performance. But finer adjustments are required for
 the PRC resonance condition, and since the cavity lengths are already fixed
 by the resonance requirements of the carrier beam (the carrier and sideband
 e-fields are relaxed in separate code executions, with the carrier first),
 the free parameter which remains to be adjusted is the sideband frequency.
 In a manner analogous to that specified for cavity length changes, sideband
 frequency adjustments for resonance are periodically computed as:
 $\Delta \nu_{\text{SB}} = \Phi \cdot ( c / 4 \pi L_{\text{PRC}})$; though
 somewhat smaller frequency changes are actually made in practice, during
 each adjustment, to lessen the disturbances to the e-field relaxation
 process. The cumulative frequency change for typical runs is small (usually
 a few hundred Hz or less), but necessary to achieve sideband PRC resonance.
 Lastly, we note that the overall results of runs for the two different
 RF-sidebands (i.e., the ``upper'' and ``lower'' sidebands,
 $\pm \nu_{\text{SB}} \equiv \nu_{\text{carr}} \pm \nu_{\text{mod}}$) are
 usually fairly similar; we generally perform computations for only one
 sideband, and assume mirror-image results for the other.

\subsubsection{\label{s3s3s3}Recycling Mirror Reflectivity Optimization}

The level of power buildup in the interferometer, and hence its GW-response,
 depends upon the reflectivities of the mirrors, and is constrained by the
 mirror losses. In LIGO, the reflectivities of most of the mirrors have been
 predetermined according a variety of auxiliary physical requirements
 (e.g., \cite{rf28}). But the principal criterion for specifying the
 reflectivity of the power recycling mirror $(R_{1})$ is the maximization of
 the gain that is achievable with power recycling. The power recycling gain,
 in turn, depends critically upon the losses experienced in the imperfect
 interferometer, which cannot be precisely determined via analytical
 estimates. $R_{1}$ is therefore a parameter that can be optimized by the
 program during each run\footnote{$R_{1}$ optimization is performed
 specifically during \textit{carrier} runs, since it is the carrier which
 requires a high PRC gain; the sidebands do not benefit from long PRC storage
 times, and would in fact suffer less degradation with a PRC storage time
 of zero, and immediate ejection at the beamsplitter exit port.}, in order
 to determine the best achievable recycling gain, given the particular
 optical imperfections being studied. The reflectivity of a real mirror,
 of course, is not something that can ordinarily be adjusted on the fly;
 rather, the results of this optimization routine in our simulation program
 have been used to help determine appropriate ``design values'' of $R_{1}$,
 for power recycling mirrors that are procured by LIGO.

For interferometer losses that are small, it can be shown (e.g., \cite{rf34})
 that the optimal choice for the recycling mirror transmission is
 $T_{\text{1,optim}} \approx A_{\text{IFO}}$, where $A_{\text{IFO}}$ is the
 effective total loss in the full interferometer (including the loss in the
 power recycling mirror). The optimal reflectivity is therefore given by:
 $R_{\text{1,optim}} = 1 - T_{\text{1,optim}}
 \approx 1 - (A_{1} - A_{\text{IFO}})$. The value of $A_{\text{IFO}}$, and
 thus of $R_{\text{1,optim}}$, will be strongly affected by optical
 deformations.

The quantity used for this optimization procedure is the \textit{real part} of
 the integrated overlap between the immediate (``prompt'') reflection of the
 laser excitation beam from the AR-side of the power recycling mirror, with
 the complete, composite e-field that is ultimately reflected back from the
 interferometer. (The \textit{imaginary} part of this overlap integral is
 related to interferometer resonance, and will essentially be zeroed if the
 cavity lengths are properly adjusted for carrier resonance in the
 interferometer, as per Sec.~\ref{s3s3s1}.) The real part of the overlap
 constitutes an error signal\footnote{This error signal is only
 \textit{precisely} valid when the power recycling mirror obeys the
 ``lossless-mirror'' Stokes condition,
 $(r_{\text{refl.}}/r_{\text{AR}}^{\ast}) = -(t/t^{\ast})$
 (cf.~Sec.~\ref{s3s1}). Our program automatically obeys the
 \textit{amplitude} part of the condition (i.e.,
 $\vert r_{\text{refl.}} \vert = \vert r_{\text{AR}}^{\ast} \vert$) for the
 recycling mirror; and though the \textit{phase} part of it is not mandated,
 we impose it for virtually all of our runs, including those described below
 in Sec.~\ref{s4}.} that can be driven to zero by changes to $R_{1}$. The
 magnitude of each change in the optimization process will be proportional
 to this error signal, though the proportionality constant (which we have
 selected through empirical tests) is not crucial, as long as it is large
 enough to achieve rapid optimization (and close convergence to the optimal
 value), while being small enough to ensure a stable optimization process.

This nulling of $\Re <\roarrow{E}_{\text{prompt}} \vert
 \roarrow{E}_{\text{total reflection}}>$ is equivalent to minimizing the
 total (``interferometer mode'') power that is reflected from the system,
 thus maximizing the power that is dissipated (and hence, that is circulating)
 inside of it. In practice, LIGO has chosen an initial value for $R_{1}$ that
 is slightly below such an ``optimized'' value, both to hedge (on the
 safer side)
 against uncertainties (and gradual increases with time) of the effective
 interferometer losses, and to provide some reflected light for length control
 signals. For the specific runs to be presented in this paper, the recycling
 mirror reflectivity has been driven all the way to $R_{\text{1,optim}}$;
 though runs can also be performed with our program in which $R_{1}$ is held
 to any particular fixed value that may be preferred due to practical
 considerations.

\subsubsection{\label{s3s3s4}Schnupp Length Asymmetry Optimization}

Also noted in Sec.~\ref{s2} was the incorporation of a macroscopic length
 asymmetry ($\sim$few tens of cm) between the inline and offline paths of the
 PRC, in order to maximally channel sideband power out through the beamsplitter
 exit port, for use as a local oscillator in the heterodyne GW-signal detection
 scheme. The maximization of this local oscillator light requires a careful
 balance between extracting sideband light from the interferometer promptly,
 before significant power is wasted due to mirror losses; yet leaving the
 sidebands in the interferometer long enough (i.e., for enough round-trip
 bounces) so that they can take full advantage of the broadband amplification
 provided by power recycling.

The total phase between the inline and offline RF-sideband fields, when they
 meet at the beamsplitter, is analytically given as: $\Phi_{\text{asymm}} =
 2 \times [-2 ~ k_{\text{mod}} ~ L_{\text{asymm}}]$, where $k_{\text{mod}}$ and
 $L_{\text{asymm}}$ are defined as in Sec.~\ref{s2}. It can be shown
 \cite{rf28} that the maximum local oscillator light is generated by the
 choice:
\begin{equation}
\cos(\Phi_{\text{asymm}}) = 2 \sqrt{R_{1} ~ R_{\text{arm avg.}}} ~
 R_{\text{BS}}~,
\label{dispeq7}
\end{equation}
where $R_{\text{arm avg.}}$ is the average reflectivity experienced by the
 beam along its inline and offline paths (including all Fabry-Perot arm
 cavity effects), and where $R_{\text{BS}} \approx T_{\text{BS}}$ for the
 beamsplitter has been assumed here\footnote{Eq.~(\ref{dispeq7}) is actually
 a simplification of the proper formula; the simulation program uses a more
 complete expression \cite{rf28}, which does not assume balance between the
 two paths or a 50\%-50\% beamsplitter, and which the program must solve
 numerically. The program also uses \textit{measured} values of
 $\Phi_{\text{asymm}}$, not this analytical one, in its
 $L_{\text{asymm}}$-optimization adjustments.}.

The reflectivity values to be used in Eq.~(\ref{dispeq7}) (or in its
 generalized version for an unbalanced interferometer) cannot be
 analytically determined before the program is
 run, since $R_{\text{arm avg.}}$ and $R_{\text{BS}}$ depend upon the losses
 due to optical imperfections, and $R_{1}$ is a variable that is optimized
 during the carrier run. The simulation program therefore uses the resonating
 powers at various interferometer locations (during the \textit{RF-sideband}
 run) to determine ``effective'' reflectivities for use in Eq.~(\ref{dispeq7}),
 to estimate the optimal choice of $\Phi_{\text{asymm}}$. Periodic changes to
 $L_{\text{asymm}}$ (totaling a few cm, cumulatively) are performed during the
 sideband field relaxation iterations, until $\Phi_{\text{asymm}}$ is brought
 to this desired value. These macroscopic length changes are applied
 antisymmetrically to the inline and offline paths, and in integral multiples
 of $\lambda_{\text{carr}} / 4$, in order to avoid (as much as possible) any
 disruption to the carrier resonance/dark-fringe conditions\footnote{We
 typically follow the initial carrier/sideband 2-run set with an additional
 carrier/sideband 2-run set, with \textit{no} changes to $L_{\text{asymm}}$
 in the latter set, to verify that the proper conditions for the carrier
 have not been disturbed.}. Lastly, we note that the program considers
 only the TEM$_{00}$ components of the e-fields while evaluating
 $\Phi_{\text{asymm}}$ and Eq.~(\ref{dispeq7}) (as opposed to considering
 the total e-fields for optimization, cf.~Sec.~\ref{s3s3s1}), since this
 extracted RF-sideband light is to be used directly for generating the
 GW-signal.

\subsubsection{\label{s3s3s5}Sideband Modulation Depth Optimization}

The sideband fields are generated via electro-optic, radio frequency modulation
 of the initial, carrier-frequency laser field. This process is represented
 mathematically as follows:
\begin{widetext}
\begin{eqnarray}
\Hat{\text{M}}\{E_{\text{las}} ~ e^{i \omega t}\} \equiv E_{\text{las}}
 ~ e^{i \omega t + i \Gamma \sin \Sigma t} = E_{\text{las}} ~
 e^{i \omega t} \cdot \sum_{n=-\infty}^{\infty} \text{J}_{n}(\Gamma)
 ~ e^{i n \Sigma t} \nonumber \\ 
 \approx E_{\text{las}} \cdot \{\text{J}_{0}(\Gamma)~e^{i \omega t} +
 \text{J}_{1}(\Gamma)~e^{i (\omega + \Sigma) t} +
 \text{J}_{-1}(\Gamma)~e^{i (\omega - \Sigma) t} \}~,
\label{dispeq8}
\end{eqnarray}
\end{widetext}
where: $E_{\text{las}}$ is the (constant) laser field amplitude,
 $\omega \equiv 2 \pi \nu_{\text{carr}}$ is the carrier angular frequency,
 $\Sigma \equiv 2 \pi \nu_{\text{mod}}$ is the modulation angular frequency,
 $\Gamma$ is the modulation depth, $\text{J}_{n}(\Gamma) = (-1)^{n}
 \text{J}_{-n}(\Gamma)$ is the Bessel function (with integer order
 \textit{n}) as a function of $\Gamma$, and where we have dropped the
 higher-order terms in the series because $\Gamma$ will be kept small enough
 for them to be unimportant.

Specifying the modulation depth is a matter of dividing up the total
 available laser power between the carrier and its RF-sidebands. The carrier
 and sidebands both contribute to the GW-signal, and due to their
 (unavoidable for the carrier, and deliberate for the sideband) ejection
 at the beamsplitter exit port, they both contribute to the shot noise that
 competes with this signal. The optimal modulation depth is determined
 via maximization of the overall ratio of GW-signal to shot noise. This
 ratio (given explicitly in Sec.~\ref{s3s4} below) is dependent upon the
 buildup of the steady-state e-fields, and hence upon the specific
 interferometer deformations which exist; for a real LIGO detector, in
 particular, the best modulation depth (or the ``best achievable"
 $\Gamma$, given technical limitations) to impose on the carrier laser
 beam can only be precisely determined once the differing performances
 of the carrier and sideband fields in the interferometer are known.

In our modeling work, the modulation depth optimization is performed as a
 post-processing step, after the simulation runs for the carrier and sideband
 e-fields have been completed. Besides the straightforward optimization of the
 shot-noise-limited GW-sensitivity (see Eq.~(\ref{dispeq11}) below) with
 respect to $\Gamma$, we are also careful to make sure of two things: that
 $\Gamma$ is small enough so that the higher-order modulation terms are
 indeed unimportant as far as their potential buildup in the LIGO
 interferometer is concerned; and that the total amount of power (carrier
 plus sidebands) falling upon the GW-signal detection photodiode at the
 beamsplitter exit port is not excessively large.

\subsubsection{\label{s3s3s6}Mirror Tilt Removal}

The final optimization procedure discussed here is a pre-processing step that
 is performed upon the mirror profile maps before they are read in during
 execution of the simulation program. Specifically, it is the removal of any
 overall \textit{tilts} due to mirror surface variations or substrate
 inhomogeneities. (A ``substrate tilt'' is merely a mirror thickness wedge,
 which can be countered in a real interferometer by a small change to the
 reference axis of a cavity.)

For irregular mirror profiles, which do not possess a unique definition of
 tilt, we choose the most useful definition for a LIGO interferometer: the
 tilt that is experienced (or ``weighted'') by a Gaussian-profile incident
 beam. To first order, the effect of mirror tilts about the two axes
 perpendicular to a (predominantly Hermite-Gaussian TEM$_{00}$) incident
 beam is to generate power in the TEM$_{10}$ and TEM$_{01}$ modes, in a
 reflection from that mirror \cite{rf9}. These modes will have imaginary
 amplitudes (assuming a real incident beam amplitude) if the mirror has no
 overall ``piston'' displacement. Given a mirror surface/substrate ``height''
 function $Z(x,y)$, and defining $M(x,y) \equiv \text{exp}[-2 i k Z(x,y)]$
 (exponentiating each pixel individually, not the whole matrix), we can remove
 the beam-weighted tilts from a given mirror deformation map by performing
 small angular corrections that set $\Im<\text{TEM}_{10} \vert M(x,y) \vert
 \text{TEM}_{00}> = \Im<\text{TEM}_{01} \vert M(x,y) \vert
 \text{TEM}_{00}> = 0$. This is done after first removing any piston offset
 from the mirror, via uniform displacements that set $\Im<\text{TEM}_{00}
 \vert M(x,y) \vert \text{TEM}_{00}>$ to zero.

\begin{figure}
\includegraphics{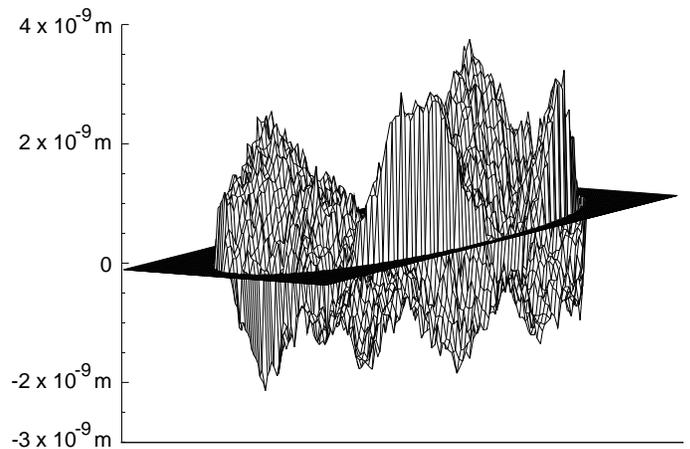}
\caption{\label{fg3}Sample map of a mirror surface with realistic deformations,
 after being processed by the tilt-removal algorithm. The mirror map, with most
 of the border region (lying beyond the finite-sized mirror apertures) clipped
 here for visual clarity, is oriented such that the incident beam propagation
 axis is along the vertical direction. The width of the grid shown here
 is $\sim 25$ cm (deformation heights not shown to scale vs.~width).}
\end{figure}

An important point about this procedure is that the appropriate beam spot
 size must be used for the modes in the overlap coefficients given above, in
 order for the ``beam-weighting'' of each mirror's tilt to be correct; but
 since the beam spot size is different at different locations in the
 interferometer, one must therefore know which specific mirror a given
 deformation map will be used for, before its tilt can be properly removed.
 Also, we note that this tilt-removal process will actually give the full
 mirror a nonzero tilt, overall; but the \textit{center} of the mirror
 (i.e., the part most sampled by the beam) will be essentially flat (on
 average) with respect to the plane transverse to the beam propagation axis.
 An example of such a tilt-removed surface deformation map is shown in
 Fig.~\ref{fg3}.

\subsection{\label{s3s4}Program Output Quantities and the GW-Sensitivity
 Function}

A full set of program runs results in a detailed specification of the final,
 steady state of the interferometer. Of the large amount of output data
 available (either directly or after some post-processing calculations), here
 are some of the most important quantities that we examine:

\textbullet ~ The \textit{relaxed powers} (for the carrier and RF-sideband
 fields) at several key locations in the interferometer.

\textbullet ~ The \textit{complete, relaxed e-fields} at all desired
 locations \textemdash ~ available for graphical display, and/or
 for complete \textit{modal decomposition} into Hermite-Gaussian
 TEM modes, which assists
 in the interpretation of interferometer conditions
 (e.g., TEM$_{10}$/TEM$_{01}$ power indicates residual tilts,
 TEM$_{20}$/TEM$_{02}$ power indicates beam mismatch, etc.).

\textbullet ~ The \textit{carrier contrast defect}, which quantifies how well
 a carrier dark-fringe was achieved at the beamsplitter exit port for the
 imperfect interferometer. Defining the carrier powers emerging from the
 relevant beamsplitter ports as $P_{\text{bright}}$ and $P_{\text{dark}}$,
 the contrast defect is given as:
\begin{equation}
1-\text{Contrast} \equiv 1-\text{C} = 1 -
 \frac{P_{\text{bright}}-P_{\text{dark}}}{P_{\text{bright}}+P_{\text{dark}}}
 \approx \frac{2 P_{\text{dark}}}{P_{\text{bright}}} ~.
\label{dispeq9}
\end{equation}
A large contrast defect implies a substantial carrier power loss at the
 beamsplitter (and thus a broadband power loss in the system), as well as a
 large carrier contribution to the shot noise at the signal port photodetector,
 and an excess of raw power falling on that photodetector.

\textbullet ~ The \textit{optimized interferometer parameters}, as described
 above in Sec.~\ref{s3s3}. In addition to their role in optimizing the
 performance of the simulated interferometer, the computed values of these
 parameters often have intrinsic importance in terms of LIGO design
 considerations.

\textbullet ~ The \textit{GW-strain-equivalent shot noise spectral density},
 $\Tilde{h}_{\text{SN}}(f)$, of a single LIGO interferometer. This crucial
 output function allows us to directly evaluate the sensitivity of the LIGO
 detector to astrophysical sources of gravitational waves, given
 interferometers with realistic optical deformations (and a realistic
 heterodyne GW-detection scheme).

\begin{figure}
\includegraphics{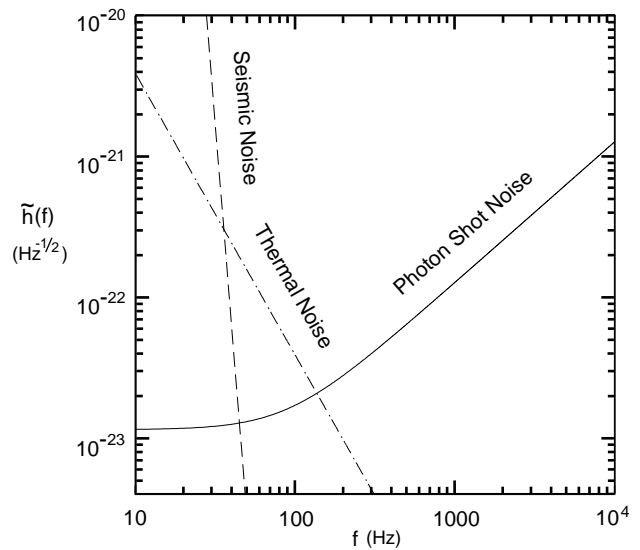}
\caption{\label{fg4}Requirement curves for the primary noise sources expected
 to limit the GW-sensitivity of first-generation LIGO interferometers.}
\end{figure}

Consider Fig.~\ref{fg4}, which shows the three most significant noise
 sources for the first-generation LIGO interferometers \cite{rf24}:
 \textit{seismic}, \textit{thermal}, and \textit{shot noise}. They are
 plotted versus GW-frequency \textit{f}, as spectral densities expressed
 in terms of the gravitational-wave Fourier amplitudes, $\Tilde{h}(f)$, that
 would induce equivalent signals in a LIGO interferometer. These particular
 curves are the first-generation LIGO requirements \cite{rf35} for the maximum
 contributions that would be acceptable from each of these three main noise
 categories\footnote{Note that the thermal noise curve in Fig.~\ref{fg4}
 actually represents an approximate conglomeration of mirror internal
 vibration noise, suspension pendulum noise, and other technical noise
 sources.}. The total noise envelope can be obtained from these individual
 curves by adding them together in quadrature (i.e., incoherent addition
 of uncorrelated noise is assumed).

Of these contributions, seismic and thermal noise are \textit{random forces}
 which will push the LIGO mirrors around in imitation of GW's. Photon
 shot noise, on the other hand, is a form of \textit{sensing noise},
 representing the quantum mechanical limit on the accuracy to which the
 mirror positions can be measured, given the finite amount of carrier power
 resonating in the FP arm cavities, and the finite amount of local oscillator
 sideband light available at the signal port. While the quality of
 interferometer optics has little effect upon the level of random force noise
 contributions (other than \textit{radiation pressure} noise, which should be
 unimportant for the first-generation LIGO interferometers \cite{rf24}), the
 quality of the optics does have a direct impact upon the sensing noise
 limitation to LIGO's GW-sensitivity. The presence of imperfect optics not
 only reduces the amount of circulating interferometer power available for
 sensing mirror positions (and thus GW-induced mirror motions), but also
 increases the amount of unwanted power at the beamsplitter
 exit port \textemdash ~ such as carrier contrast defect power, and
 RF-sideband power in non-TEM$_{00}$ modes \textemdash ~ which contribute
 to the shot noise, but not to the GW-signal. We
 therefore focus upon the \textit{shot-noise-limited} region of the LIGO noise
 envelope in evaluating the effects of optical imperfections. For each set of
 output results, $\Tilde{h}_{\text{SN}}(f)$ can be computed, and compared
 either to the first-generation LIGO requirements (see Sec.~\ref{s4} below),
 or to astrophysical predictions (see Sec.~\ref{s5}), in order to determine
 the effects of optical deformations upon LIGO's ability to detect GW's of
 reasonable, anticipated strengths.

A full derivation of the formula for $\Tilde{h}_{\text{SN}}(f)$ is given
 elsewhere \cite{rf28}. Here we present the resulting expression, in terms
 of the relevant output data from our simulation program. The shot noise
 sensitivity limit is expressed in terms of the strength of the GW needed
 to produce a signal that could match this noise. For a monochromatic
 gravitational wave of the form \cite{rf6}:
\begin{equation}
h^{\text{TT}}(f)= \sqrt{2} ~ h \cos(2 \pi f t + \phi_0)~,
\label{dispeq10}
\end{equation}
which is incident upon a single interferometric detector with optimal incidence
 angle and polarization, a GW-amplitude of $h = \Tilde{h}_{\text{SN}}(f)$ would
 produce a unity signal-to-shot-noise ratio when sampled over unity bandwidth
 (i.e., 1 second integration time). With these definitions, we have:
\begin{widetext}
\begin{eqnarray}
\Tilde{h}_{\text{SN}}(f) \equiv \{\frac{\text{Signal}(f)/h}{
 \text{Noise}}\}^{\text{-1}} = 
 \{\sqrt{\eta ~ \nu_{\text{carr}} ~ P_{\text{las}} ~ / ~ h_{\text{pl}}}
 ~~ \text{J}_{0}(\Gamma) ~ \text{J}_{1}(\Gamma)
 ~ \frac{4 \sqrt{2} \pi \tau_{\text{s}} \cdot r_{\text{FP arm back}} 
 \cdot t_{\text{FP arm input}}}{\sqrt{1+(4 \pi \tau_{\text{s}} f)^{2}}}
 \nonumber \\
 \times \frac{\sqrt{P^{00}_{\text{SB, exit port}}} ~ (r_{\text{BS}} \cdot 
 \sqrt{P^{00}_{\text{carr, inline FP-arm}}} + t_{\text{BS}} \cdot 
 \sqrt{P^{00}_{\text{carr, offline FP-arm}}})}{\sqrt{\text{J}_{0}(\Gamma)^{2}
 ~ P^{\text{tot}}_{\text{carr, exit port}} + 3 ~ \text{J}_{1}(\Gamma)^{2}
 ~ P^{\text{tot}}_{\text{SB, exit port}} }} \}^{-1} ~,
\label{dispeq11}
\end{eqnarray}
\end{widetext}
where: $P_{\text{las}}$ is the total excitation laser power (before
 radio frequency modulation) in Watts, $h_{\text{pl}}$ is Planck's
 constant, $\nu_{\text{carr}}$ is the carrier frequency, $\eta$ is
 the quantum efficiency of the photodetector at the signal port,
 $\text{J}_{0}(\Gamma)$ and $\text{J}_{1}(\Gamma)$ represent the
 division of laser power between the carrier and either one of its
 RF-sidebands (cf.~Eq.~(\ref{dispeq8})), and terms like
 ``$P^{\text{tot}}_{\text{carr, exit port}}$'', etc., represent
 the \textit{dimensionless} relaxed power value (in either the
 TEM$_{00}$ mode or the total in all modes, as indicated) that is
 reported by the numerical simulation code for a carrier or sideband
 e-field in the indicated interferometer location. These
 ``dimensionless'' power values for the simulated e-fields are all
 normalized in the program to an input carrier/sideband laser beam
 power of 1 Watt. Note that we have not included both the upper and
 lower sidebands separately here; in many cases it is sufficient
 to plug the simulation results for either of them into
 Eq.~(\ref{dispeq11}), and assume the interferometer performance
 to be fairly symmetrical about the carrier frequency for these
 RF-modulated fields.

Some remaining quantities in Eq.~(\ref{dispeq11})
 (e.g., $r_{\text{FP arm back}}$, $t_{\text{BS}}$) are estimated mirror
 (amplitude) reflection and transmission coefficients for the path that any
 GW-induced signal fields would take from the FP-arms (where they are
 physically generated) through the interferometer, until they exit at the
 beamsplitter signal port. Finally, the quantity $\tau_{\text{s}}$ is the
 \textit{effective storage time} of the GW-induced signal fields in the
 realistically-deformed FP arm cavities. Since the explicit simulation of the
 buildup of these signal e-fields in the arm cavities would involve an
 additional set of e-field relaxation procedures that are not performed by
 the program, the effective storage time of these GW-induced e-fields in the
 (imperfect) arm cavities must be approximated, as follows:

\begin{widetext}
\begin{equation}
\tau_\text{s} \approx \frac{L}{c} ~ \cdot ~ \frac{1}{\sqrt{P^{00}_{
\text{carr, Recyc. Cav.}}}} ~ \cdot ~ \\
\frac{1}{2} \{\frac{\sqrt{P^{00}_{\text{carr, inline FP-arm}}}}{t_{\text{BS}}
 \cdot t_{\text{FP arm input}}} + \frac{\sqrt{P^{00}_{
 \text{carr, offline FP-arm}}}}{r_{\text{BS}} \cdot t_{\text{FP arm input}}}\}
 ~,
\label{dispeq12}
\end{equation}
\end{widetext}
where $L$ is the length of the FP arm cavities, $c$ is the speed of light, and
 a properly-weighted average has been performed over the results for the inline
 and offline FP-arms. We note that the dependence of $\Tilde{h}_{\text{SN}}(f)$
 upon the GW-frequency is contained within the expression
 $\sqrt{1+(4 \pi \tau_{\text{s}} f)^{2}}$, so that it has two basic regimes
 separated by the ``knee'' or pole frequency, $f_{\text{pole}} =
 1/(4 \pi \tau_{\text{s}})$, as follows:
\begin{equation}
\label{dispeq13}
\Tilde{h}_{\text{SN}}(f)\vert_{f \ll f_{\text{pole}}} \approx \text{constant} ~, 
~~ \Tilde{h}_{\text{SN}}(f)\vert_{f \gg f_{\text{pole}}} \propto f ~.
\end{equation}
These two regimes are evident in the plot of $\Tilde{h}_{\text{SN}}(f)$ that is
 included in Fig.~\ref{fg4}.

\section{\label{s4}A Selection of Results Obtained with the Simulation Program}

In this section, we will demonstrate the code with a set of runs incorporating
 measurement maps made from very high quality mirrors. The only interferometer
 imperfections included are these mirror deformation maps, the finite sizes of
 the mirrors, and a small amount of pure loss specified for each mirror. The
 pre-specified mirror loss values represent absorption in the mirror substrates
 and coatings, as well as high-angle scattering due to roughness finer than the
 resolution of our grids (for which most of the scattered power is lost beyond
 the finite mirror apertures, particularly in the long FP-arms). For each
 individual run in this set, the simulated interferometer is completely
 optimized in the sense of Sec.~\ref{s3s3}. The program input parameters
 common to all runs are those listed in Table \ref{tbl1}.

First, however, we note that many tests of the simulation program have been
 performed \cite{rf36} to ensure its validity as a realistic model of a LIGO
 interferometer. These tests include: comparisons with numerical simulation
 results in the literature \cite{rf16,rf37}, to verify basic operations like
 beam propagation and diffractive loss from finite mirrors; comparisons against
 modal analysis methods \cite{rf9} for simple interferometer imperfections,
 such as mirror tilts; and comparisons against analytical methods \cite{rf38}
 for computing the effects of Zernike polynomial \cite{rf39} mirror surface
 deformations. We have implemented anti-aliasing and energy-conservation
 procedures (cf.~Sec.~\ref{s3s1}) to make sure that the physics of the
 interferometer is being properly simulated; and we have performed numerous
 ``common-sense'' tests to check that the program's output results not only
 make physical sense, but also reasonably reflect the types of interferometer
 imperfections being modeled. Finally, direct comparisons of our simulation
 results with experimental measurements for large-scale interferometers are
 now becoming possible (e.g., \cite{rf21,rf22}), and have thus begun to provide
 useful mutual feedback for both experimental and modeling efforts.

\subsection{\label{s4s1}The Realistically-Deformed Mirror Maps Used for these
 Runs}

The mirror maps used in these simulation runs have been derived from two
 measurements of real optical components: the first one, obtained by LIGO
 from Hughes-Danbury Optical Systems, is a phase map of the reflection from
 the polished surface of the ``Calflat'' reference flat mirror used by the
 AXAF program (e.g., \cite{rf40}) for the calibration of their extremely
 smooth, high-resolution conical mirrors; and the second one is a transmission
 phase map of a trial LIGO mirror substrate obtained from Corning. Both of
 these measurements were of uncoated, fused silica substrates. Measurement
 maps of fully-coated mirrors were not available for the runs in this study.

Each of these near-LIGO-quality mirror maps (one surface reflection map and
 one substrate transmission map) were extrapolated into an array of many maps,
 so that we could place deformed surfaces and substrates upon all of the
 interferometer mirrors simultaneously. One of us (Y.~H.) used a three-step
 process for creating new mirror deformation maps: first, the surface (or
 substrate) source map was Fourier transformed into spatial-frequency space;
 then, the relative phases of its Fourier components were randomized; and
 finally, the application of an inverse Fourier transform created a new mirror
 with randomized features, but with the same power spectrum of
 deformations\footnote{Although this process does not preserve
 \textit{coherent} structures \textemdash ~ e.g., spikes, etc.,
 which might be systematic effects of mirror fabrication
 \textemdash ~ such structures were manually added in by us
 for other tests, enabling us to place limits on their significance.}. This
 process was performed many times, to create 15 new surface maps out of the
 initial Calflat map, and 7 new substrate maps out of the initial Corning map.
 Test runs with various groupings of these mirrors have demonstrated to us
 that different members of a family of randomized mirrors have (as expected)
 very similar characteristics to one another in terms of their effects upon
 interferometer performance, and that swapping one for another has little
 effect upon the output results of the simulation program.

Further preparation steps for the mirror maps were taken to adapt them to the
 appropriate grid parameters and mirror aperture dimensions
 (cf.~Table \ref{tbl1}), and also to supply related deformation maps for the
 beamsplitter, given its $45^{\circ}$ tilt angle and consequently elliptical
 apertures; full details of all such preparations are given in \cite{rf28}.
 Each of the resulting mirror maps were then tilt-removed (using appropriate
 beam spot sizes) with respect to normal-incidence laser beams, as per
 Sec.~\ref{s3s3s6}. An example of one of these surface deformation maps has
 been shown in Fig.~\ref{fg3}.

As a final step, it was necessary to create several families of surface maps
 with different levels of deformations, in order to help the LIGO Project
 evaluate a range of mirror polishing specifications for the procurement
 of the core optics, as well as to make room for the not-fully-determined
 effects of mirror coating deformations. To accomplish this, the
 Calflat-derived surface deformation height maps were uniformly multiplied
 by scale factors to generate the new families. The original family of
 surfaces \textemdash ~ which possess RMS deformations
 of $\sim$.6 nm when sampled over their central 8 cm
 diameters \textemdash ~ has been labeled ``$\lambda / 1800$''
 (with $\lambda \equiv \lambda_{\text{YAG}} = 1.064 ~ \mu \text{m}$). The
 scaled-up families are labeled $\lambda / 1200$, $\lambda / 800$, and
 $\lambda / 400$, respectively. The mirror substrate maps (possessing RMS
 deformations of $\sim$1.2 nm over their central 8 cm diameters), however,
 were considered likely to represent the best quality of fused silica
 substrates obtainable for the first-generation LIGO interferometers
 \cite{rf41}, and were not re-scaled; all of the deformed-mirror runs
 discussed below were done with this same family of substrate maps. Note
 that a direct comparison of the substrate RMS value with that of the surfaces
 is not informative, since the substrates are typically sampled less frequently
 by the electric fields (particularly for the carrier light resonating in the
 FP arm cavities), and are therefore less significant (for the carrier fields,
 at least) in their effects.

Given these families of mirror deformation maps, it becomes possible to
 examine the overall performance of a coupled-cavity LIGO interferometer
 in the presence of ``realistic" mirror deformations, to comprehensively
 estimate its true capabilities.

\subsection{\label{s4s2}The Results of the Runs}

This study contains results from five separate simulation runs: one run with
 perfectly smooth mirror substrates and surfaces, and four runs with:
 (i) deformed substrate maps for all of the mirrors, plus, (ii) deformed
 surface maps for all mirrors from, respectively, the $\lambda / 1800$,
 $\lambda / 1200$, $\lambda / 800$, or $\lambda / 400$ families. For all
 cases other than the ``perfect mirrors'' run, the transmission and
 (reflective-side) reflection maps for each 2-port mirror were constructed
 from one surface phase map and one substrate phase map; the program then
 derives the AR-side reflection map from the other maps via energy
 conservation (i.e., the lossless-mirror Stokes condition,
 cf.~Sec.~\ref{s3s1}). The beamsplitter's two reflection and two
 transmission operators were constructed from one surface map and two
 substrate maps, with the remaining map derived via its generalized energy
 conservation formula.

The results are summarized in Table \ref{tbl2}. Several of the quantities
 described in Sections \ref{s3s3} and \ref{s3s4} are included, as well as the
 true (``absolute'') carrier and sideband power exiting at the beamsplitter
 signal port. The DC values and pole frequencies of the GW-sensitivity curves
 calculated for each run are also given, from which one can construct
 $\Tilde{h}_{\text{SN}}(f)$ for each case as follows
 (cf.~Eq's.~(\ref{dispeq11})-(\ref{dispeq13})):
\begin{equation}
\Tilde{h}_{\text{SN}}(f) \equiv \Tilde{h}_{\text{SN}}(0) \cdot
 \sqrt{1+(f/f_{\text{pole}})^{2}} ~.
\label{dispeq14}
\end{equation}
Some of these quantities have also been re-computed in the hypothetical
 case of an \textit{idealized output mode cleaner} functioning at the
 signal port, which would act to strip away all of the non-TEM$_{00}$
 light (contributing only to noise) from the exiting beams, while passing
 all of the TEM$_{00}$ light (contributing all of the signal and some
 shot noise) through to the output photodetector. A photodetector quantum
 efficiency of $\eta = 0.8$ was assumed for computing the values of
 $\Tilde{h}_{\text{SN}}(f)$, $\Gamma$, etc.

\begingroup
\squeezetable
\begin{table*}
\caption{\label{tbl2}Output results for the series of interferometer
 simulation runs performed using realistic deformation maps for the optical
 surfaces and substrates. A total (pre-modulation) laser input power of
 6 Watts is assumed (except where otherwise noted), as well as a
 photodetector quantum efficiency of $\eta = 0.8$.}
\begin{ruledtabular}
\begin{tabular}{llllll}
\textbf{Quantity} & \multicolumn{5}{c}{\textbf{Values For Specified Run}}\\
\hline
\hline
Deformed Surfaces (RMS in wavelengths) & Zero & $\lambda_{\text{YAG}} /
 1800$ & $\lambda_{\text{YAG}} / 1200$ & $\lambda_{\text{YAG}} / 800$ &
 $\lambda_{\text{YAG}} / 400$\\
Deformed Substrates (Y/N) & No & Yes & Yes & Yes & Yes\\
Recycling Mirror Reflectivity\footnotemark[1] & 98.61\% & 98.37\% &
 98.07\% & 97.39\% & 93.90\% \\
Schnupp Length Asymmetry (cm)\footnotemark[1] & 9.0 & 12.3 & 13.5 & 15.9 &
 24.7\\
TEM$_{00}$ Carrier Power, Recycling Cavity\footnotemark[2] & 72.40 & 61.54 &
 51.84 & 38.41 & 16.37\\
TEM$_{00}$ Carrier Power, Fabry-Perot Arm Avg.\footnotemark[2] & 4726.7 &
 4012.0 & 3374.3 & 2491.7 & 1042.4\\
TEM$_{00}$ Carrier Power, Exit Port\footnotemark[2] & $2.70 \times 10^{-6}$ &
 $9.29 \times 10^{-6}$ & $1.94 \times 10^{-5}$ & $4.56 \times 10^{-5}$ &
  $1.89 \times 10^{-4}$\\
Total Carrier Power, Exit Port\footnotemark[2] & $2.15 \times 10^{-3}$ &
 $8.53 \times 10^{-3}$ & $1.43 \times 10^{-2}$ & $2.25 \times 10^{-2}$ &
  $3.65 \times 10^{-2}$\\
Carrier Contrast Defect, $1-\text{C}$ & $6.02 \times 10^{-5}$ & $2.82
 \times 10^{-4}$ & $5.62 \times 10^{-4}$ & $1.20 \times 10^{-3}$ & $4.73
  \times 10^{-3}$\\
TEM$_{00}$ 1-Sideband Power, Recycling Cavity\footnotemark[2] & 59.08 &
 28.32 & 25.01 & 20.17 & 10.66\\
TEM$_{00}$ 1-Sideband Power, Exit Port\footnotemark[2] & .9067 & .6745 &
 .6955 & .7344 & .8196\\
Total 1-Sideband Power, Exit Port\footnotemark[2] & .9071 & .7590 & .7761 &
 .8082 & .8795\\
GW-Response Pole Frequency, $f_{\text{pole}}$ (Hz) & 90.32 & 90.38 & 90.45 &
 90.61 & 91.45\\
Modulation Depth, $\Gamma$ \footnotemark[1] & 0.279 & 0.405 & 0.455 & 0.501 &
 0.549\\
Absolute Carrier Exit-Port Power (mW) & 12.41 & 47.1 & 77.2 & 118.6 & 187.8\\
Absolute 2-Sideband Exit-Port Power (mW) & 207.1 & 358.9 & 458.3 & 571.3 &
 737.7\\
DC GW-Sensitivity, $\Tilde{h}_{\text{SN}}(0)$ \footnotemark[1] & $4.79
 \times 10^{-24}$ & $5.76 \times 10^{-24}$ & $6.41 \times 10^{-24}$ &
  $7.59 \times 10^{-24}$ & $1.20 \times 10^{-23}$\\
\hline
\hline
\multicolumn{6}{c}{\textbf{Re-Computed Values Assuming Ideal Output Mode
 Cleaner:}}\\
\hline
\hline
Modulation Depth, $\Gamma$ \footnotemark[1] & 0.053 & 0.078 & 0.093 & 0.113 &
 0.156\\
Absolute Carrier Exit-Port Power (mW) & $1.6 \times 10^{-3}$ & $5.6
 \times 10^{-3}$ & 0.12 & 0.27 & 1.10\\
Absolute 2-Sideband Exit-Port Power (mW) & 7.7 & 12.2 & 17.9 & 28.1 & 59.6\\
DC GW-Sensitivity, $\Tilde{h}_{\text{SN}}(0)$ \footnotemark[1] & $4.61
 \times 10^{-24}$ & $5.01 \times 10^{-24}$ & $5.48 \times 10^{-24}$ &
  $6.40 \times 10^{-24}$ & $1.00 \times 10^{-23}$ \\
\end{tabular}
\end{ruledtabular}
\footnotetext[1]{Denotes parameter optimized by program, or during
 post-processing.}
\footnotetext[2]{Denotes quantity normalized to 1 Watt of
 carrier/sideband excitation light power.}
\end{table*}
\endgroup

\begin{figure}
\includegraphics{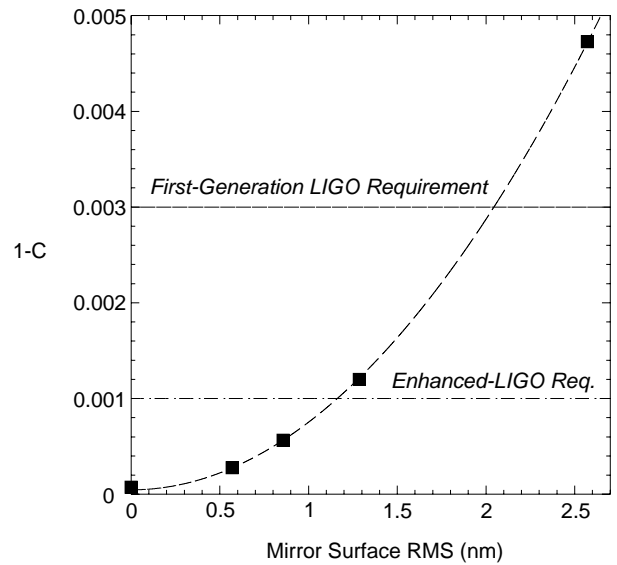}
\caption{\label{fg5}The carrier contrast defect, $1-\text{C}$, plotted
 versus RMS mirror surface deformations. The dashed line is a quadratic
 fit to the points representing the runs performed with, respectively:
 ``perfect mirrors'', then the $\lambda / 1800$, $\lambda / 1200$,
 $\lambda / 800$, and $\lambda / 400$ mirrors. The horizontal lines are
 upper limits on the contrast defect values allowed for the first-generation
 LIGO and enhanced-LIGO interferometers, respectively.}
\end{figure}

Utilizing this output data, one noteworthy result is that several effects
 of deformed optics have a \textit{quadratic} dependence upon the
 deformation amplitudes\footnote{We note that the signal-generating
 sideband power at the beamsplitter exit port does \textit{not} worsen
 quadratically with RMS deformation levels; and in fact,
 Table \ref{tbl2} shows an \textit{increase} in exit-port sideband power, 
 as the deformation levels get very large. This counterintuitive behavior
 is due to great effectiveness of the parameter optimizations
 (cf.~Sec.~\ref{s3s3}) \textemdash ~ specifically that of the Schnupp
 length asymmetry, which is forced to a larger value for highly deformed
 mirrors, in order to get the sideband fields out of the degraded
 interferometer as soon as possible. This type of behavior for the
 sideband fields in the presence of highly deformed mirrors, though
 technically correct, may be unduly optimistic when considered in the
 context of a real LIGO system, in which the many parameters are not as
 easily adjustable \textemdash ~ if at all adjustable \textemdash ~ in
 the experimental system, as they are in the simulation program.}.
 This is as expected, since power scattered out of
 a Gaussian beam by mirror roughness scales like the square of the roughness
 amplitude, even when the deformations are spread over a range of spatial
 frequencies \cite{rf42}. For example, the effects of imperfect mirrors on
 the FP arm cavity power buildup can be expressed in terms of an equivalent
 ``effective mirror loss'' that would analytically reproduce the same amount
 of circulating FP-arm power. Doing this for each of the runs, we obtain an
 effective loss function that increases quadratically (versus mirror surface
 RMS deformation amplitude) from the baseline value of $\sim$50 parts per
 million of ``absorptive'' loss that is put in by hand for each mirror
 \cite{rf28}. Similarly, the contrast defect ($1-\text{C}$), which comes
 from power coupled into non-TEM$_{00}$ beam modes by (longer spatial
 wavelength) optical imperfections, is also well represented by a quadratic
 fit, as is shown in Fig.~\ref{fg5}. Lastly, the \textit{optimized} recycling
 mirror (power) reflectivity, $R_{1}$, should decrease quadratically from
 its ``perfect mirrors'' value, since 1-$R_{\text{1,optim}}$ is directly
 proportional to interferometer losses (cf.~Sec.~\ref{s3s3s3}), and the
 dominant losses (contrast defect loss and high-angle scattering in the
 FP-arms) are both quadratically dependent upon mirror deformation RMS.
 A fit of $R_{\text{1,optim}}$ vs. RMS \cite{rf28} does indeed bear out this
 expected functional form (though not all the way to a surface RMS of zero,
 since other effects then take over \textemdash ~ substrate
 deformations, absorptive losses, etc.).

\begin{figure}
\includegraphics{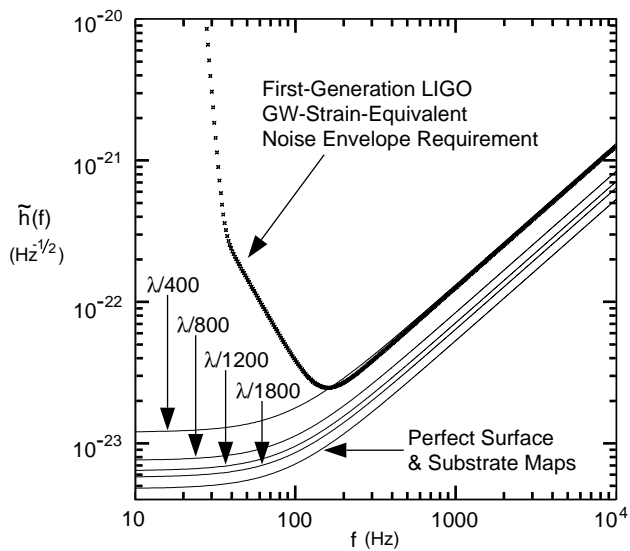}
\caption{\label{fg6}Comparison of the shot noise curves,
 $\Tilde{h}_{\text{SN}}(f)$, computed for each of the interferometer
 simulation runs, against the official GW-strain-equivalent noise envelope
 requirement for the first-generation LIGO interferometers.}
\end{figure}

The most significant issue to address was whether LIGO would be able to
 perform according to Project requirements \cite{rf43}, given mirrors with
 realistic levels of optical deformations. That question is answered in the
 affirmative, as demonstrated in Table \ref{tbl2} and Fig's.
 \ref{fg5}-\ref{fg6}, for all cases except that of the worst surfaces simulated
 here. First of all, the first-generation LIGO interferometers are required
 to have a carrier power gain of at least 30 in the PRC, and this target is
 achieved in all runs except for the (ultra-conservative) run performed with
 $\lambda / 400$ surface deformations. In addition, the first-generation
 interferometers must have a contrast defect of $1-\text{C} < 3
 \times 10^{-3}$, a requirement that is also satisfied by all runs other then
 the $\lambda / 400$ case (and anything better than $\sim \lambda / 500$
 would suffice). Furthermore, it has been quoted \cite{rf44} that the
 ``enhanced'' LIGO interferometers should satisfy the more stringent
 requirement of $1-\text{C} < 1 \times 10^{-3}$, which would be achieved
 by three of the five simulation runs here (and anything better than
 $\sim \lambda / 900$ would suffice) \textemdash ~ an acceptable
 result, especially considering the likelihood of improved mirror
 quality by the time the enhanced
 interferometers are operational. One caveat, however, is that although the
 contrast defect requirements are met for most of the runs, a large amount of 
 total power (several hundred milliwatts) falls upon the output photodetector
 in all cases, especially for runs with highly deformed mirrors in which the
 optimized sideband modulation depth is large, in order to help the signal
 compete against increased shot noise. If a signal detection apparatus that
 could handle this large amount of power cannot be supplied, then an output
 mode cleaner may be needed; Table \ref{tbl2} shows that an output mode
 cleaner would greatly reduce the power that the photodetector must accommodate
 (while also improving the GW-sensitivity by $\sim$15\%), as long as it
 operates closely enough to an ``idealized'' performance, as described above.

The most fundamental requirement to be satisfied is that the
 shot-noise-limited sensitivity curve for an interferometer with deformed
 mirrors (cf.~Eq's.~(\ref{dispeq11}), (\ref{dispeq14})) should fall within
 the bounds of the \textit{GW-strain-equivalent noise envelope requirement},
 which is the official total-noise limit set for the (full 4 km baseline)
 first-generation LIGO interferometers \cite{rf43}. In Fig.~\ref{fg6}, the
 computed $\Tilde{h}_{\text{SN}}(f)$ curves for these runs are plotted against
 the data points that define the LIGO requirement envelope, with all data given
 in terms of spectral densities. The seismic, thermal, and shot-noise-dominated
 regimes of the requirement envelope are apparent in the figure, and our
 $\Tilde{h}_{\text{SN}}(f)$ curves can be compared to the shot-noise-dominated
 region. We conclude, once again, that all runs other than the $\lambda / 400$
 case succeed in meeting the initial-LIGO requirement. The overall conformity
 of these results indicates that there is a very clear \textemdash ~ and
 very strict, though achievable \textemdash ~ quality level for the core
 optics that must be reached
 in order for the first-generation LIGO interferometers to achieve their target
 performances.

\section{\label{s5}Impact of Optical Deformations upon LIGO Science
 Capabilities}

To place the results of our runs into a scientifically relevant perspective,
 we estimate the effects of optical deformations upon LIGO's ability to detect
 gravitational waves from anticipated astrophysical events. Focusing here upon
 GW-sources that might be detectable by the first-generation LIGO
 interferometers, and considering cases for which an improvement in the shot
 noise limit may significantly increase the number of detection events (or
 enlarge the detectable range of some reasonably well-understood scientific
 parameter), we arrive at two good candidates for study: \textit{periodic} GW's
 from non-axisymmetric pulsars, and \textit{burst} GW's from the coalescence
 of black hole/black hole (BH/BH) binaries.

\subsection{\label{s5s1}Non-Axisymmetric Pulsars}

To facilitate comparisons against theoretical signal estimates, the output
 data from the simulation program are used to create representative
 interferometer noise curves. To that end, we form the quadrature sum
 (for each run) of three functions: the seismic and thermal noise
 \textit{requirement} curves (from Fig.~\ref{fg4}), added to the particular
 shot noise curve, $\Tilde{h}_{\text{SN}}(f)$, that is computed
 (cf.~Eq.~(\ref{dispeq11}), with data from Table \ref{tbl2}) for each of
 the simulation runs.

These summed, spectral density noise curves must be converted into useful
 signal-to-noise expressions. We follow the conventions of Thorne \cite{rf6},
 in which each of these total-noise curves ($\Tilde{h}_{\text{SN}}(f)$) is
 converted to a dimensionless expression ($h_{3/\text{yr}}$), and is compared
 to the ``characteristic strength'' ($h_{\text{c}}$) of a GW-source. For
 periodic sources, the condition $h_{\text{c}} = h_{3/\text{yr}}$ means that
 after \textit{coincidence detection} in two identical interferometers for
 one-third of a year of integration time, a source with strength
 $h_{\text{c}}$ can be extracted from the Gaussian noise with a confidence
 level of 90\%.

Averaging over all polarizations and orientations of the source on the sky,
 and treating the GW-frequency $f$ (equal to twice the pulsar's rotation
 frequency) and phase as known, Eq.~52a of \cite{rf6} yields:
\begin{equation}
h_{3/\text{yr}} \vert_{f} \approx 1.7 \cdot \sqrt{5} \cdot \sqrt{10^{-7}
 \text{Hz}} \times \Tilde{h}(f) ~.
\label{dispeq15}
\end{equation}

For a pulsar with \textit{gravitational ellipticity} $\varepsilon$, a
 rotation-axis moment of inertia of $10^{45} ~ \text{g} \cdot \text{cm}^2$,
 radiating at frequency $f$ at a distance $r$ from the earth, and averaging
 over orientation angles of the source, we have (from Eq.~55 of \cite{rf6}):
\begin{equation}
h_{\text{c}} \vert_{f} \approx 7.7 \times 10^{-20} \cdot \varepsilon
 \cdot (\frac{10 ~ \text{kpc}}{r}) \cdot (\frac{f}{1 ~ \text{kHz}})^{2} ~.
\label{dispeq16}
\end{equation}

\begin{figure}
\includegraphics{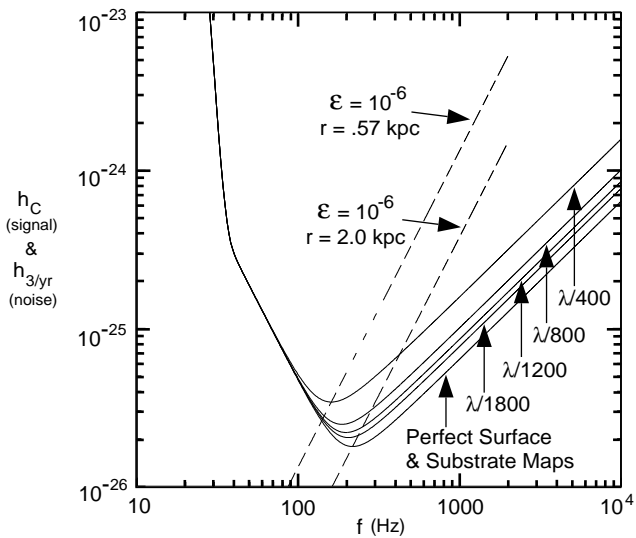}
\caption{\label{fg7}Plots of characteristic GW-signal strength
 $h_{\text{c}}$ versus frequency, for pulsars with specified ellipticity
 and distance from the earth (dashed lines), displayed against the
 dimensionless noise curves $h_{3/\text{yr}}$ (for periodic searches) that
 are computed from the output results of the simulation runs (solid lines).}
\end{figure}

In Fig.~\ref{fg7}, $h_{3/\text{yr}}$ is plotted for all of the simulation
 runs\footnote{This figure differs slightly from the corresponding one
 (Fig.~2) in \cite{rf45}, because a somewhat older formulation (Eq's.~12
 and 13 of \cite{rf46}, each with factor of 2 corrections) was used there
 for the thermal noise requirement curve. The numbers quoted in the ensuing
 discussion here differ slightly as a consequence.}, and displayed with them
 are two $h_{\text{c}}$ curves, each representing the \textit{locus} of
 possible GW-strengths (plotted versus frequency, up to $f \approx 2 ~ \text {kHz}$) for 
a pulsar with given $\varepsilon$ and $r$. We have chosen
 $\varepsilon = 10^{-6}$, which should be below the ``breaking strain'' of
 neutron star crusts \cite{rf47}, yet may produce a detectable signal. While
 this value is too large for millisecond pulsars given typical limits on their
 rates of GW-induced spin-down \cite{rf47}, it may not be unreasonable for
 newly-formed pulsars, of which it has been estimated \cite{rf48} that there
 may be $\sim$25 such ``new'' pulsars in the galaxy, or one every $\sim$few kpc.

By setting $h_{\text{c}} = h_{3/\text{yr}}$ at the frequency of peak
 sensitivity ($f_{\text{peak}}$) for a given $h_{3/\text{yr}}$ curve, one
 obtains the rough estimate that in going from the $\lambda / 400$ case to
 (or very nearly to) the perfect mirrors case, the typical ``lookout distance''
 to which such a pulsar is detectable increases from $\sim$.57 kpc to 2.0 kpc
 (while $f_{\text{peak}}$ increases to $\sim$200 Hz). This improvement roughly
 increases the expected number of pulsar detections (in a galactic disk
 distribution) by $\sim (2.0/.57)^{2} \approx 12$, and brings the lookout
 distance for the detection of even a \textit{single} $\varepsilon = 10^{-6}$
 pulsar with the first-generation LIGO interferometers to a more reasonable
 value. Alternatively, for a pulsar at a given distance and GW-emission
 frequency, it provides a factor of $\sim (2.0/.57) \approx 3.5$ leeway
 in the smallest detectable value of $\varepsilon$.

\subsection{\label{s5s2}Black Hole/Black Hole Binary Coalescences}

A similar formulation is used for the analysis of burst sources. Each
 quadrature noise sum, $\Tilde{h}(f)$, is again computed; and with
 appropriate angle averaging and assumptions for optimal filtering,
 we have (from Eq.~34 of \cite{rf6}):
\begin{equation}
h_{3/\text{yr}} \vert_{f} \approx \sqrt{5} \cdot \sqrt{\ln(f/10^{-8} 
~ \text{Hz})} \cdot \sqrt{f} \times \Tilde{h}(f) ~.
\label{dispeq17}
\end{equation}
For bursts, $h_{\text{c}} = h_{3/\text{yr}}$ means that after coincidence
 detection in two identical interferometers for one-third of a year of
 observation time, a detection of GW-strength $h_{\text{c}}$ has a
 90\% probability of being a real signal, rather than an accidental
 conspiracy of the Gaussian noise in the detectors. (Coincidence operation
 between multiple interferometers should theoretically eliminate the false
 burst signals caused by non-Gaussian, uncorrelated noise \cite{rf24}, which
 we neglect here.)

We consider a BH/BH binary system with equal component masses,
 $M_{1} = M_{2} = 10 M_{\odot}$, located a distance $r$ from the earth,
 and evolving in frequency through the inspiral phase until it reaches the
 onset of the coalescence phase (i.e., merger and ringdown) at
 $f_{\text{merger}} \sim 205 ~ \text{Hz} \cdot (20 M_{\odot} / M_{\text{tot}})
 \approx 200 ~ \text{Hz}$ \cite{rf49}. For these parameters, Eq.~46b of
 \cite{rf6} yields (with a cumulative factor of 2 adjustment from factor
 of 2 corrections \cite{rf50} to Eq's.~29 and 44 of \cite{rf6}):
\begin{equation}
h_{\text{c}} \vert_{f=f_{\text{peak}}} \approx 5.0 \times 10^{-21}
 \cdot (\frac{100 ~ \text{Mpc}}{r}) \cdot (\frac{100 ~ \text{Hz}}{f})^{1/6} ~.
\label{dispeq18}
\end{equation}
The proper way to interpret this formula, is that \textit{if} the frequency
 of peak detector sensitivity is $f \equiv f_{\text{peak}}$, then the
 \textit{total} integrated inspiral signal deposited into the detector
 (for comparison with $h_{3/\text{yr}}$) is determined by $h_{\text{c}}$
 evaluated specifically at that $f_{\text{peak}}$.

\begin{figure}
\includegraphics{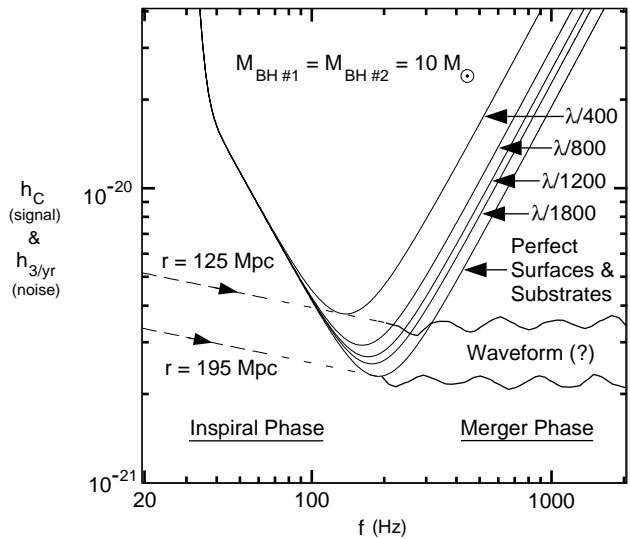}
\caption{\label{fg8}Plots of characteristic GW-signal strength
 $h_{\text{c}}$ as a function of the detector's peak sensitivity frequency,
 during the inspiral phase of 10 $\text{M}_{\odot}$ black hole/black hole
 binaries (dashed lines), displayed against the dimensionless noise curves
 $h_{3/\text{yr}}$ (for burst searches) that are computed from the output
 results of the simulation runs (solid lines).}
\end{figure}

In Fig.~\ref{fg8}, we have plotted the $h_{3/\text{yr}}$ noise curves
 (for burst searches) that are obtained for each of the simulation runs,
 along with two $h_{\text{c}}$ curves (with arrows showing time evolution),
 for coalescence events that just manage to skirt the high-detection-confidence
 threshold of $h_{\text{c}} = h_{3/\text{yr}}$ at $f_{\text{peak}}$ during the
 course of their inspirals. The first conclusion that one may draw from this
 plot is that LIGO's sensitivity to these coalescence events (at least during
 inspiral) is most strongly limited by the \textit{low frequency} part of the
 noise curves, i.e., the seismic and thermal noise limits. Nevertheless, there
 is a measurable benefit from improving the shot noise limit: judging from the
 plot, it can be estimated that in going from the $\lambda / 400$ case to the
 perfect mirrors case, the lookout distance is increased from $\sim$125 Mpc
 to $\sim$195 Mpc; or equivalently, it increases the expected number of
 detectable events by the factor $\sim (195/125)^{3} \approx 3.8$. The actual
 rate of BH/BH coalescence events is extremely uncertain (even their existence
 is uncertain), but good middle-of-the-road values that one could use as
 benchmarks are the ``best estimates'' that have been made by Phinney
 \cite{rf51}, and Narayan \textit{et al.} \cite{rf52}, which are,
 respectively: $\sim$3 per year out to 200 Mpc (assuming a Hubble constant
 of $H_{0} \approx 75 ~ \text{km} ~ \text{s}^{-1} ~ \text{Mpc}^{-1}$), and
 $\sim$1 per year out to $200 ~ \text{Mpc} \times (100 ~ \text{km} ~
 \text{s}^{-1} ~ \text{Mpc}^{-1} / H_{0})$. Thus, the perfect mirrors run
 appears to put BH/BH binary coalescence events just within the conceivable
 reach of detection for the first-generation LIGO interferometers. Perhaps
 even more significantly, improving the shot noise limit could increase LIGO's
 sensitivity to the onset of the merger phase of BH/BH binaries with masses
 like these; GW-emission during the actual merger is still poorly understood,
 but it may involve the most powerful radiation of detectable energy during
 the overall coalescence process \cite{rf49}.

We end this section by cautioning that the aforementioned numbers must be
 considered very rough estimates, given the highly simplified ways in which
 we have treated the noise curves of real interferometers, the sophisticated
 data analysis methods needed to extract signals from the noise, and the many
 inherent uncertainties about the GW-sources themselves.
 In fact, rather than interpreting the results of this section as
 indicating how LIGO \textit{optics} would determine initial-LIGO
 \textit{physics}, it would be more appropriate to interpret these results
 as a demonstration of how initial-LIGO \textit{physics} places
 requirements on the \textit{optics}. The firm scientific conclusion that
 one can draw, however, is that using optical components of the best
 achievable quality can indeed make a difference in whether or not the
 first-generation LIGO interferometers have a fighting chance to detect
 gravitational waves from these promising astrophysical sources.

\section{\label{s6}Discussion and Conclusions}

We have shown our simulation program to be useful for gaining physical
 insight into interferometer behavior, and for roughly estimating the
 effects of optical deformations upon LIGO science capabilities. Due
 to the highly detailed nature of the model, it has been an effective tool
 for research and development in the LIGO Project.
 To date, it has been used to address
 several important design issues\footnote{We are grateful to E. D'Ambrosio,
 P. Fritschel, K. Ganezer, B. Kells, N. Mavalvala, R. Savage, D. Shoemaker,
 D. Sigg, K. Sliwa, A. Weinstein, J. Weldon, S. Whitcomb, and H. Yamamoto for
 their help in the application of our simulation work to these and other
 LIGO-related issues.}, providing support for technical initiatives such
 as: (i) Aiding LIGO in its transition from Argon-ion lasers
 ($\lambda = 514.5 ~ \text{nm}$) to Nd:YAG lasers ($\lambda = 1.064 ~ \mu
 \text{m}$) for the main carrier beam \cite{rf53}, and demonstrating that
 interferometer performance is less sensitive to mirror figure deformations
 (of a specified physical amplitude) when larger-wavelength laser beams are
 used; (ii) Assisting in the selection of the Schnupp Length Asymmetry scheme
 for GW-signal readout over an alternative, external modulation (Mach-Zehnder)
 scheme \cite{rf54}; (iii) Helping model the performances (particularly the
 effects of optical deformations upon interferometer control systems) of the
 major LIGO prototype interferometers, including the Fixed Mass Interferometer
 (FMI) \cite{rf54} and the Phase Noise Interferometer (PNI) \cite{rf55} at MIT,
 and the 40-meter interferometer at Caltech \cite{rf56}; (iv) Providing
 assistance in the selection of optical parameters for the long-baseline LIGO
 interferometers, such as beam spot sizes, mirror curvatures, and aperture
 sizes (particularly the perspective-foreshortened beamsplitter aperture)
 \cite{rf57}; (v) Conducting preliminary studies of the usefulness of an output
 mode cleaner at the beamsplitter signal port (cf.~Sec.~\ref{s4s2}); (vi)
 Simulating the effects of refraction index variations due to \textit{thermal
 lensing} (e.g., \cite{rf58}) on interferometer performance, resulting in
 preliminary estimates of $\sim$15\% degradation to $\Tilde{h}_{\text{SN}}(f)$
 from mirror coating absorption values of 0.6 ppm, or (equivalently) from
 substrate bulk absorption values (in fused silica) of
 $5 ~ \text{ppm}/\text{cm}$ \cite{rf59}.

Perhaps the most important use of the program has been its involvement
 in the LIGO ``Pathfinder Project'' \cite{rf60}, the initiative to set
 specifications and tolerances for LIGO's core optical components, and to
 procure them through a cooperative effort of several vendors and optics
 metrology groups. Our program has also been used in conjunction with other
 modeling initiatives at LIGO \cite{rf11,rf12,rf61}, to create a broad-based
 interferometer simulation environment involving different algorithmic
 approaches and physical regimes of interest.

The latest versions of our program continue to be used to address important
 questions raised by the LIGO Project, such as estimating the performance
 of advanced-LIGO detectors \cite{rf30,rf56,newref1} \textemdash ~ i.e.,
 interferometers incorporating Dual Recycling \cite{rf23}, or even
 Resonant Sideband Extraction \cite{rf58} \textemdash ~ in the
 presence of optical deformations, and participating
 in initiatives to set core optical specifications (and to design the control
 systems) for those advanced detectors \cite{rf30,rf62}. Many related issues
 will undoubtedly arise in the near future, as advanced interferometer
 configurations (and increasingly better optics) become available, for which
 this program can be used as a primary modeling tool for LIGO and its
 collaborating gravitational wave groups.

\begin{acknowledgments}
We are grateful to Jean-Yves Vinet and Patrice Hello of the VIRGO Project
 for supplying us with the original code that formed the early basis of our
 work; and to Hughes-Danbury and Corning for their generosity and spirit of
 research in sharing their data with us. We would like to thank D. Shoemaker
 and D. Sigg for their help in the early preparation of this manuscript.
 The development of our simulation program was principally supported by NSF
 Cooperative Agreement PHY-9210038.
\end{acknowledgments}

\end{document}